\documentclass[]{aa}
\usepackage{graphicx,natbib}

\def\aV{\mbox{$\rm A_V$}}

\def\jh{\mbox{$(J-H)$}}

\def\jk{\mbox{$(J-K_s)$}}

\def\ebv{\mbox{$E(B-V)$}}

\def\ejh{\mbox{$E(J-H)$}}

\def\rc{\mbox{$R_{\rm c}$}}

\def\rl{\mbox{$R_{\rm RDP}$}}
\def\rx{\mbox{$R_{\rm ext}$}}

\def\ds{\mbox{$d_\odot$}}

\def\jj{\mbox{$J$}}
\def\hh{\mbox{$H$}}
\def\ks{\mbox{$K_s$}}

\begin{document}

\title{Star clusters or asterisms? 2MASS CMD and structural analyses of 15 challenging targets}

\author{E. Bica\inst{1} \and C. Bonatto\inst{1}}

\offprints{C. Bonatto}

\institute{Universidade Federal do Rio Grande do Sul, Departamento de Astronomia\\
CP\,15051, RS, Porto Alegre 91501-970, Brazil\\
\email{charles@if.ufrgs.br, bica@if.ufrgs.br}
\mail{charles@if.ufrgs.br} }

\date{Received --; accepted --}

\abstract
{Poorly-populated star clusters may have photometric and structural properties 
not much different from asterisms, to the point that, in some cases, widely-used 
databases present conflicting classifications.}
{We investigate the nature of a sample of challenging targets that have been classified 
either as star clusters or asterisms in different studies. A few objects are studied 
for the first time.}
{The analysis employs 2MASS photometry, field-star decontamination, to enhance 
the intrinsic colour-magnitude diagram (CMD) morphology, and colour-magnitude 
filters, for high contrast stellar radial density profiles (RDPs).}
{Based on properties derived from field-star decontaminated CMDs, and structural 
parameters from RDPs, we find that Pismis\,12, IC\,1434, Juchert\,10, Ruprecht\,30, 
NGC\,3519, Herschel\,1, Mayer\,1, and Muzzio\,1 are open clusters with ages within 
5\,Myr - 1.3\,Gyr. Ruprecht\,129, 130, 140, and 146 are borderline cases, being 
rather poorly-populated, with evolutionary sequences and RDPs suggesting star clusters. 
Dolidze\,39, BH\,79, and Ruprecht\,103, have CMDs and RDPs typical of asterisms.}
{When a low stellar population is associated with a dense field contamination and/or 
important differential reddening, only a thin line separates star clusters and 
asterisms. These cases require specific analytical tools to establish their nature. }


\titlerunning{Star cluster or asterism: case studies}

\maketitle

\section{Introduction}
\label{Intro}

Establishing the cluster nature of a stellar overdensity requires meeting 
criteria stemming from two different perspectives, one related to the 
spatial distribution of the stars, and the other to the colour-magnitude 
diagram (CMD) morphology. In short, a star cluster's overdensity should 
be spatially extended and present a CMD morphology significantly different 
from that of the field stars. Otherwise, it may simply be a statistical 
fluctuation of the field or a low-absorption window, a so-called asterism. 

Usually, a star cluster's overdensity distributes radially from its centre
following an analytical profile, such as the single-mass, isothermal spheres
of \citet{King1962} and \citet{King1966}, the modified isothermal sphere of 
\citet{Wilson75}, or the power-law with a core of \citet{EFF87}. However, this
expectation does not necessarily apply to clusters in all scales. Faint and/or
poorly-populated star clusters may have a stellar radial density profile (RDP) 
that does not follow any analytical profile (e.g. \citealt{FSR20}). Instead, 
the very poorly-populated ones may have an RDP characterised essentially by a 
central overdensity together with significant noise outwards, typical features
of an asterism (e.g. Fig.~9 of \citealt{FSR20}). In these cases, only the CMD 
morphology may provide clues to the overdensity nature.

With respect to the photometric properties, the Galactic open clusters (OCs) 
vary a lot in CMD morphology, such as the presence of evolutionary sequences 
ranging from a $\sim1$\,Myr to a few Gyr (e.g. \citealt{OvrlTeut}; \citealt{Rup15}), 
including here the very young embedded (in gas and dust) clusters (ECs), which have 
CMDs dominated by pre-main sequence (PMS) stars (e.g. \citealt{Orto08}). These 
clusters can also present significant differences in terms of observable stellar 
magnitude ranges, the amount of stellar fore/background contamination, as well as 
foreground and internal differential reddening (especially the ECs). All these 
aspects should be taken into account when identifying and characterising a star 
cluster. 

Several approaches have been employed to cope with these difficulties, adapted to 
the available photometric range, but usually in the optical or near infrared (e.g.
\citealt{OvrlTeut}; \citealt{Monteiro10}; \citealt{Alessi03}; \citealt{Subra10}). 
For instance, it is common to find works based on {\em (i)} extracted CMDs without
applying field star decontamination (e.g. \citealt{Orto05}), {\em (ii)} proper
motion-filtering (e.g. \citealt{Dias06}) or, {\em (iii)} CMD field decontamination 
(e.g. \citealt{OvrlTeut}). Historically, optical cluster parameters were the main 
source of information, but in recent years the near-infrared, using mostly the 
uniform 2MASS\footnote{The Two Micron All Sky Survey, All Sky data release 
(\citealt{2mass1997}) - {\em http://www.ipac.caltech.edu/2mass/releases/allsky/}}, has 
become a major data input, especially for reddened clusters. Our group, in particular, 
has been employing 2MASS CMDs and RDPs built with statistical field decontamination 
and colour-magnitude filters (\citealt{BB07}). We have studied clusters in different 
evolutionary stages, since EC formation in complexes (\citealt{Sh2-132}) to their 
eventual dissolution (e.g. \citealt{Bochum1}; \citealt{Pi5}; \citealt{vdB92}). The 
respective photometric and structural parameters, have been shown to constrain the 
study of particular targets. 

Different targets, in general, require analyses with particular approaches, until 
getting a clear picture of their nature. However, when the targets are poorly-populated, 
which applies to most Galactic OCs and ECs, the star cluster characterisation would 
further require: {\em (i)} a cluster nature determination for the first time by means 
of derivation of fundamental and structural parameters, and {\em (ii)} two or more 
studies on the same object to allow for parameter comparisons and consistency checks. 

OC parameters are assembled in two comprehensive and frequently updated databases, 
WEBDA\footnote{\em www.univie.ac.at/webda} and the ``Catalog of Optically Visible Open 
Clusters and Candidates'' (DAML02) \footnote{\em www.astro.iag.usp.br/~wilton}. In some 
cases, objects have conflicting classifications in both databases. In this respect, 
\citet{PN06} inspected the status of OC 
parameters among different studies, finding 72 standard ones with well-determined and 
consistent values. Their analysis invokes the need of repeated cluster studies to assess 
the external error of a given parameter. Thus, it is important to derive accurate parameters 
for newly studied clusters and revisit ambiguous ones with different approaches. Such an 
effort, in turn, would constantly replenish the OC databases with updated parameters, 
which is fundamental to obtain external errors. This also would allow assessing OC 
statistical properties (e.g. \citealt{DiskProp}; \citealt{Piskunov09}), and help 
deepen studies related to star cluster formation and evolution (e.g. \citealt{PZ10}).

In the present study we focus on 15 challenging stellar concentrations, in general very 
field-contaminated  and/or absorbed, and either indicated as OC or asterism in previous 
studies, if any. We analyse the targets with 2MASS data and the tools that we have developed
to explore the field-star 
distribution and/or reddening variations in the area, in view of finding further evidence to 
confirm or discard any cluster. We propose that such borderline objects can be used as a 
measure of ``cluster nature reproduceability'' among different studies, which may be taken 
as a counterpart to that of \citet{PN06}. 

\begin{table*}
\caption[]{General data on the targets}
\label{tab1}
\tiny
\renewcommand{\tabcolsep}{0.4mm}
\renewcommand{\arraystretch}{1.25}
\begin{tabular}{lcccccccccccc}
\hline\hline
&\multicolumn{6}{c}{Literature or improved from DSS/XDSS images}&&\multicolumn{5}{c}{This work}\\
\cline{2-7}\cline{9-13}\\
Name&$\ell$&$b$&$\alpha(2000)$&$\delta(2000)$&D&Other Designations&&$\alpha(2000)$&$\delta(2000)$&Age&\aV&\ds\\
    &(\degr)&(\degr)&(hms)&($\degr\,\arcmin\,\arcsec$)&(\arcmin)& &&(hms)&($\degr\,\arcmin\,\arcsec$)&(Myr)&(mag)&(kpc)\\
    (1)&(2)&(3)&(4)&(5)&(6)&(7)&&(8)&(9)&10)&(11)&(12)\\
\hline
\multicolumn{13}{c}{Open clusters}\\
\hline
IC\,1434      & 99.93 &$-$2.67 & 22:10:22 &$+$52:51:30 &   6 & Mel\,239, Cr\,445, OCl-223&&22:10:33.4 &$+$52:50:33.7 &$800\pm100$&$0.4\pm0.1$&$2.6\pm0.6$\\
Mayer\,1      &119.46 &$-$0.92 & 00:22:03 &$+$61:44:57 &   6 & OCl-288.1&& 00:21:54.0 &$+$61:45:32.0 &$\sim10$&$2.0\pm0.5$&$2.2\pm0.5$\\
Herschel\,1   &219.35 &$+$12.35& 07:47:02 &$+$00:01:06 &  33 & ADS\,6366, ASCC\,41, FSR\,1141&& 07:47:02.0 &$+$00:01:17.2 &$300\pm100$&$0.4\pm0.3$&$0.35\pm0.05$\\
Ruprecht\,30  &246.42 &$-$4.04 & 07:42:25 &$-$31:28:21 &   6 & OCl-687, ESO\,429SC15&& 07:42:26.5 &$-$31:27:54.7 &$400\pm100$&$1.2\pm0.5$&$5.4\pm1.3$\\
Muzzio\,1     &267.96 &$-$1.30 & 08:58:04 &$-$47:41:20 &  11 & FSR\,1456&& 08:58:03.2 &$-$47:42:16.5 &$\sim5$&$3.2\pm0.5$&$1.3\pm0.3$\\
Pismis\,12    &268.63 &$+$3.21 & 09:20:00 &$-$45:06:55 &   7 & OCl-765, BH\,62, ESO\,261SC5&& 09:20:00.3 &$-$45:06:54.2 &$1300\pm300$&$1.5\pm0.3$&$1.9\pm0.3$\\
NGC\,3519     &290.43 &$-$1.12 & 11:04:07 &$-$61:22:06 &   5 & Ru\,93, OCl-844, ESO\,128SC30&& 11:04:09.6 &$-$61:22:30.0 &$400\pm100$&$1.0\pm0.3$&$2.0\pm0.5$\\
Juchert\,10   &316.00 &$-$0.29 & 14:40:16 &$-$60:22:20 &   4 & GLP-53, FSR\,1686, SAI\,121&& 14:40:18.6&$-$60:22:51.5 &$800\pm200$&$6.4\pm0.6$&$3.2\pm0.8$\\
\hline
\multicolumn{13}{c}{Borderline cases}\\
\hline
Ruprecht\,140 &  0.04 &$-$8.85 & 18:21:51 &$-$33:12:38 &   4 & OCl-1, ESO\,395SC1&& 18:21:56.2 &$-$33:12:32.0 &$4000\pm1000$&$0.0\pm0.3$&$3.5\pm0.8$\\
Ruprecht\,146 & 14.10 &$-$9.66 & 18:52:31 &$-$21:04:55 &   4 & OCl-39, ESO\,592SC4&& 18:52:30.8 &$-$21:04:41.5 &$1000\pm300$&$1.9\pm0.3$&$1.9\pm0.4$\\
Ruprecht\,130 &359.23 &$-$0.95 & 17:47:33 &$-$30:05:28 & 4.5 & OCl-1034, BH\,247, ESO\,455SC41&& 17:47:32.9 &$-$30:05:12.2 &$100\pm50$&$7.3\pm0.3$&$0.9\pm0.2$\\
Ruprecht\,129 &359.59 &$-$0.64 & 17:47:12 &$-$29:37:10 & 8.5 & OCl-1037, ESO\,455SC40&& 17:47:11.5 &$-$29:35:24.0 &$500\pm100$&$3.2\pm0.3$&$2.7\pm0.6$\\
\hline
\multicolumn{13}{c}{Probable asterisms}\\
\hline
Dolidze\,39   & 75.51 &$+$1.52 & 20:16:04 &$+$37:53:27 &  15 & OCl-159&& 20:16:04.0 &$+$37:53:27.0 &\\
BH\,79        &277.01 &$-$0.10 & 09:43:41 &$-$53:15:00 &   8 & && 09:43:43.7 &$-$53:14:50.2 &\\
Ruprecht\,103 &298.28 &$+$4.13 & 12:16:57 &$-$58:25:35 &   4 & OCl-876, ESO\,130SC11&& 12:16:57.8 &$-$58:25:47.2 &\\
\hline
\end{tabular}
\begin{list}{Table Notes.}
\item Col.~6: diameter estimated from CADC Sky Survey images; Col.~12: distance 
from the Sun. Cluster acronyms: Mel = Mellote, Cr = Collinder, GLP = GLIMPSE.
\end{list}
\end{table*}

This paper is organised as follows. In Sect.~\ref{RecAdd} we present literature data
on the present object sample. In Sect.~\ref{2mass} we discuss the 2MASS photometry 
and build the field-star decontaminated CMDs. In Sect.~\ref{DFP} we derive fundamental 
parameters. In Sect.~\ref{struc} we estimate structural parameters. In Sect.~\ref{Conclu} 
we discuss the nature of the objects according to our analyses, and provide our concluding 
remarks.

\section{Previous Results}
\label{RecAdd}

In Table~\ref{tab1} we condense identification data for the targets, arranged according to 
the final classification (Sect.~\ref{Conclu}). Most of the equatorial and Galactic coordinates, 
as well as angular diameters, were optimised from those in the original catalogues, by means 
of visual inspections of digitised sky survey DSS and XDSS images\footnote{Extracted from 
the Canadian Astronomy Data Centre (CADC), at \em http://cadcwww.dao.nrc.ca/}. The data in
Table~\ref{tab1} are important as input values for the subsequent analyses, since exceedingly 
large offsets may affect the convergence of the results. DSS or XDSS images of the targets are 
shown in App.~\ref{appA}. In App.~\ref{appB} we show near-infrared images of the reddened
clusters Mayer\,1, Muzzio\,1, and Juchert\,10. Below we discuss what the literature contains 
about the selected targets.

\subsection{Seven Ruprecht targets}
\label{RuOCs}

Ruprecht objects tend to be relatively faint and field contaminated. They are listed 
in the classical early catalogue containing OCs and candidates by \citet{Alter70}.
Recently, \citet{Rup15} studied 15 such neglected Ruprecht clusters using the present 
method to determine their fundamental and structural parameters. We now extend the 
analysis to 7 additional Ruprecht objects (hereafter Ru) with dubious classifications, 
in view of constraining their properties.

\citet{PC01b} studied Ru\,103, 129, and 146 with BVI photometry concluding that the 
CMDs do not contain clear evolutionary sequences, and stellar counts do not not show any  
significant contrast with respect to the background field. Likewise, \citet{PC01a} did 
not find any meaningful CMD sequence in the optical range for Ru\,140. They concluded that 
these objects are asterisms, which would be consistent with the rich disk and/or bulge
fields against which they are projected (Table~\ref{tab1}).

The coordinates of the stellar concentration NGC\,3519 (\citealt{Sinnott1988}) have Ru\,93 
as counterpart in the OC catalogue of \citet{Alter70}. The object is also present in the 
ESO/Uppsala catalogue (\citealt{Lauberts82}). \citet{Kharch05a} derived the distance from
the Sun $\ds=1.4$\,kpc and the age $\approx440$\,Myr for NGC\,3519. However, \citet{CSB10}, 
using optical photometry concluded that no cluster was located in the area.

\citet{PCB00} studied Ru\,130 by means of BVI photometry and optical integrated spectra, 
and concluded it to be an OC at $\ds=2.1$\,kpc and $50\pm10$\,Myr of age. \citet{PNIM06}
found $\ds=1.8$\,kpc and $80\pm20$\,Myr of age. Similarly to Ru\,129, Ru\,130 has a 
remarkable background field, which is consistent with the angular separations of only 0.76\degr\ 
and 1.2\degr, respectively from the Galactic centre. The populous and very absorbed central 
bulge is challenging for any analysis method. The present near-infrared tools, with emphasis 
on field star decontamination, are particularly performing in such cases, especially with  
the use of  colour-magnitude filters (e.g. \citealt{FSR20}).

\citet{VMC08} derived $\ds=1.3\pm0.2$\,kpc and $60\pm15$\,Myr of age for Ru\,30. However,
\citet{Carraro2010} studied the stellar field in the direction of Ru\,30 concluding that
no cluster is present. 

\subsection{Three targets from the sample of \citet{MN08}}
\label{MN08}

\citet{MN08} combined BV and 2MASS photometry with proper motions to analyse doubtful 
OCs. They concluded that Dolidze\,39, IC\,1434, and Mayer\,1 were composed of physically 
unrelated stars. However, by means of proper motions, DAML02 finds that Dolidze\,39 is
an OC at $\ds=1.4$\,kpc. \citet{Tad2009} employed 2MASS photometry without field 
decontamination and indicated that IC\,1434 is an OC at $\ds=3.0$\,kpc and with $320$\,Myr 
of age. \citet{Kharch05a}, using 2MASS data and proper motions, concluded that Mayer\,1 
is a cluster of 55\,Myr of age, located at 1.4\,kpc from the Sun, and reddening $\ebv=0.4$. 

We note that Dolidze\,39 is apparently related to a nebula. It is projected on the 
Cygnus OB\,1 Association. We will check in particular if PMS stars occur in the area,
which would be consistent with the young age. Mayer\,1 has faint extended nebulosities 
in the area, both in B and R CADC images. Again, PMS stars may be present.

\subsection{Additional targets}
\label{AddTgt}

Herschel\,1 was found by John Herschel (\citealt{H27}) and described as a triple star 
in a small cluster. This stellar concentration was probably not included in the later 
catalogues of non-stellar objects because of his focus at the time on the double star 
catalogue. The triple star was also included in the double star catalogue of \citet{Aitken32} 
as the triple ADS\,6366. SIMBAD\footnote{\em http://simbad.u-strasbg.fr/simbad/} lists 
the ADS members as HD\,63065 (spectral type B9), BD+00\,2080 (A2), and BD+00\,2079C 
(F5IV). \citet{Kharch05b} concluded it to be an OC with 275\,Myr of age, located at 
370\,pc from the Sun, and reddening $\ebv=0.02$. The object is also included in the 
FSR survey (\citealt{Froeb07}). Herschel\,1 has a detached core, which makes it an 
interesting target for a detailed structural analysis.

Juchert\,10 was reported in the optical by M. Juchert. A private communication of  
M. Kronberger to DAML02 in 2005 indicated it as an asterism. The object was as well
detected as a GLIMPSE cluster in the far infrared (\citealt{Mercer05}). GLIMPSE 
clusters are also referred to as Mercer and [MCM2005b] clusters (e.g. \citealt{Hanson10}; 
\citealt{Kurtev07}). The object has recently been detected in the FSR (\citealt{Froeb07}) 
and  SAI (\citealt{Glushkova10}) surveys. The latter authors found an age of 
$\approx900$\,Myr, $E(B-V)=1.48$, and $\ds=4.02$\,kpc.

\citet{Muzzio79} communicated Muzzio\,1 as a new OC, possibly related to RCW\,38 in 
Vela. From UBV photometry, he derived a distance of $\ds=1.7$\,kpc for Muzzio\,1.

The cluster Pismis\,12 was found in the survey of \citet{Pismis59}. Despite being
relatively bright (App.~\ref{appA}), it has not been studied so far (WEBDA; DAML02).

BH\,79 was first reported by \citet{vdBH75}. They communicated a cluster seen in red 
plates, but not so clearly in the blue. It was concluded to be poor with an angular 
diameter of 2.5\arcmin. We point out that a faint stellar concentration (Table~\ref{tab1}) 
is very close to the original coordinates, and we probe its nature.

\section{Photometric analysis}
\label{2mass}

2MASS provides an all-sky coverage, together with spatial and photometric uniformity.
These properties are essential for wide angular extractions that, in turn, can be
used for  statistically characterising the colour and magnitude distribution of the 
field stars (Sect.~\ref{2mass}) and determining the background level (Sect.~\ref{struc}). 
Thus, to meet these criteria, the photometry for each target was extracted from 
VizieR\footnote{\em http://vizier.u-strasbg.fr/viz-bin/VizieR?-source=II/246} in a wide 
circular field of radius $\rx=60-90\arcmin$, varying according to each case. As a photometric 
quality constraint, only stars with \jj, \hh, and 
\ks\ errors lower than 0.1\,mag are used. Reddening corrections are based on the absorption 
relations $A_J/A_V=0.276$, $A_H/A_V=0.176$, $A_{K_S}/A_V=0.118$, and $A_J=2.76\times\ejh$ 
given by \citet{DSB2002}, with $R_V=3.1$, considering the extinction curve of \citet{Cardelli89}. 

The first step for correctly characterising the photometric and structural properties of a star
cluster is to find its actual centre, i.e., the maximum of the overdensity produced by member 
stars. To do this, we {\em (i)} download the photometry centred on the input coordinates
(cols.~4 and 5 of Table~\ref{tab1}), {\em (ii)} build the RDP (Sect.~\ref{struc}) for deriving 
a first estimate of target size ($R=\rl$) and location of the comparison field ($R_{FS1}\la 
R\la R_{FS2}$), {\em (iii)} build the decontaminated (Sect.~\ref{DecontCMD}) CMD of the region 
$R\la\rl$, {\em (iv)} re-compute the central coordinates\footnote{In the present approach, the 
centre corresponds to the coordinates that produce the smoothest stellar RDP and, at the same 
time, the highest density in the innermost region.}, now with the decontaminated photometry, 
and {\em (v)} repeat steps {\em (i) - (iii)} for the new central coordinates. Usually, there 
are small differences between our coordinates (cols.~8 and 9 of Table~\ref{tab1}) and the 
input ones. The efficiency of this approach can be assessed by the shape of the resulting 
RDPs (Fig.~\ref{fig5}. All cases present a rather high central density dropping relatively 
smoothly outwards.

\begin{figure}
\resizebox{\hsize}{!}{\includegraphics{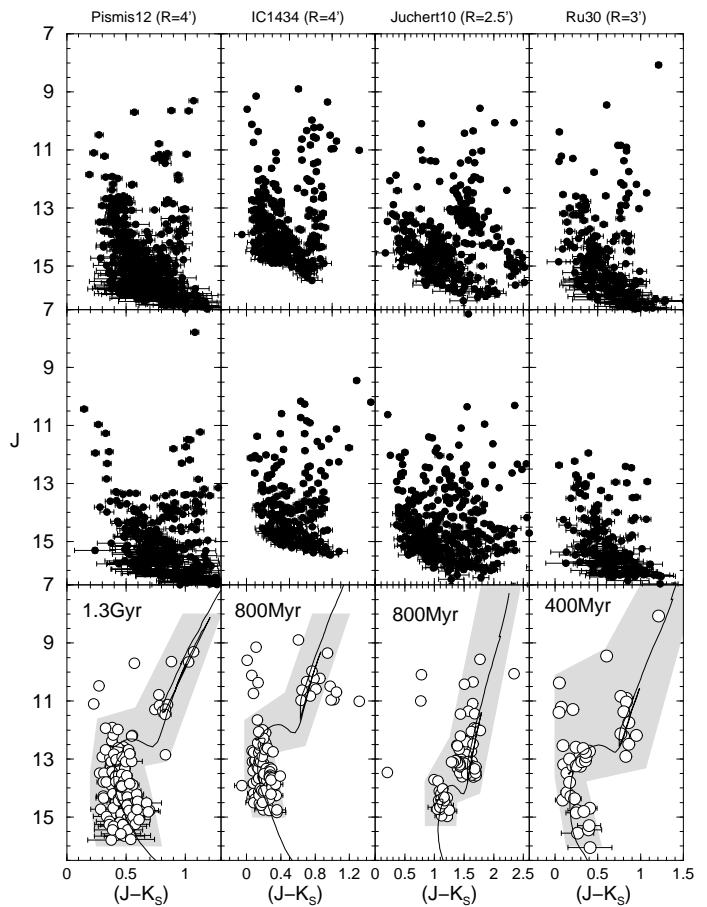}}
\caption[]{CMDs of Pismis\,12, IC\,1434, Juchert\,10, and Ru\,30, for
a representative cluster region (top panels) and the equal-area comparison field 
(middle). The decontaminated CMDs (bottom) are shown together with the isochrone 
solution (solid line) and the colour-magnitude filter (shaded polygon). Note that, 
for most stars, the error bars are smaller than the symbol.}
\label{fig1}
\end{figure}

\begin{figure}
\resizebox{\hsize}{!}{\includegraphics{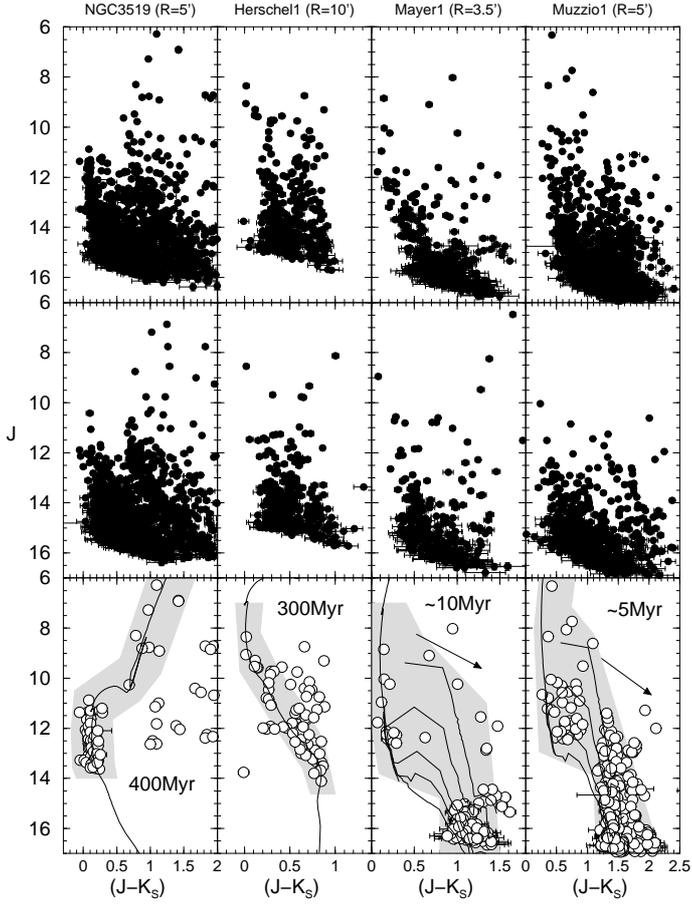}}
\caption[]{Same as Fig.~\ref{fig1} for NGC\,3519, Herschel\,1, Mayer\,1, and Muzzio\,1.
PMS tracks correspond to the ages (from top to bottom) 0.2, 1, 5, and 10\,Myr in
Muzzio\,1, reaching to 20\,Myr in Mayer\,1. A reddening vector for $\aV=0$ to $\aV=5$ 
is shown in the CMDs of Mayer\,1 and Muzzio\,1.}
\label{fig2}
\end{figure}

\begin{figure}
\resizebox{\hsize}{!}{\includegraphics{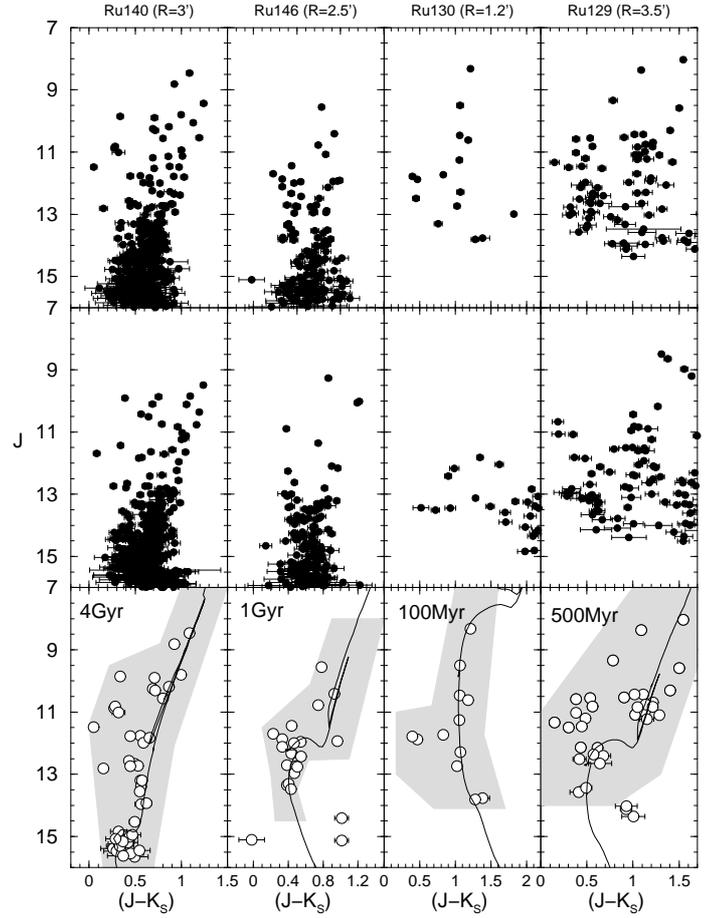}}
\caption[]{Same as Fig.~\ref{fig1} for the limiting cases Ru\,140, 146, 130, and 129.}
\label{fig3}
\end{figure}

\begin{figure}
\resizebox{\hsize}{!}{\includegraphics{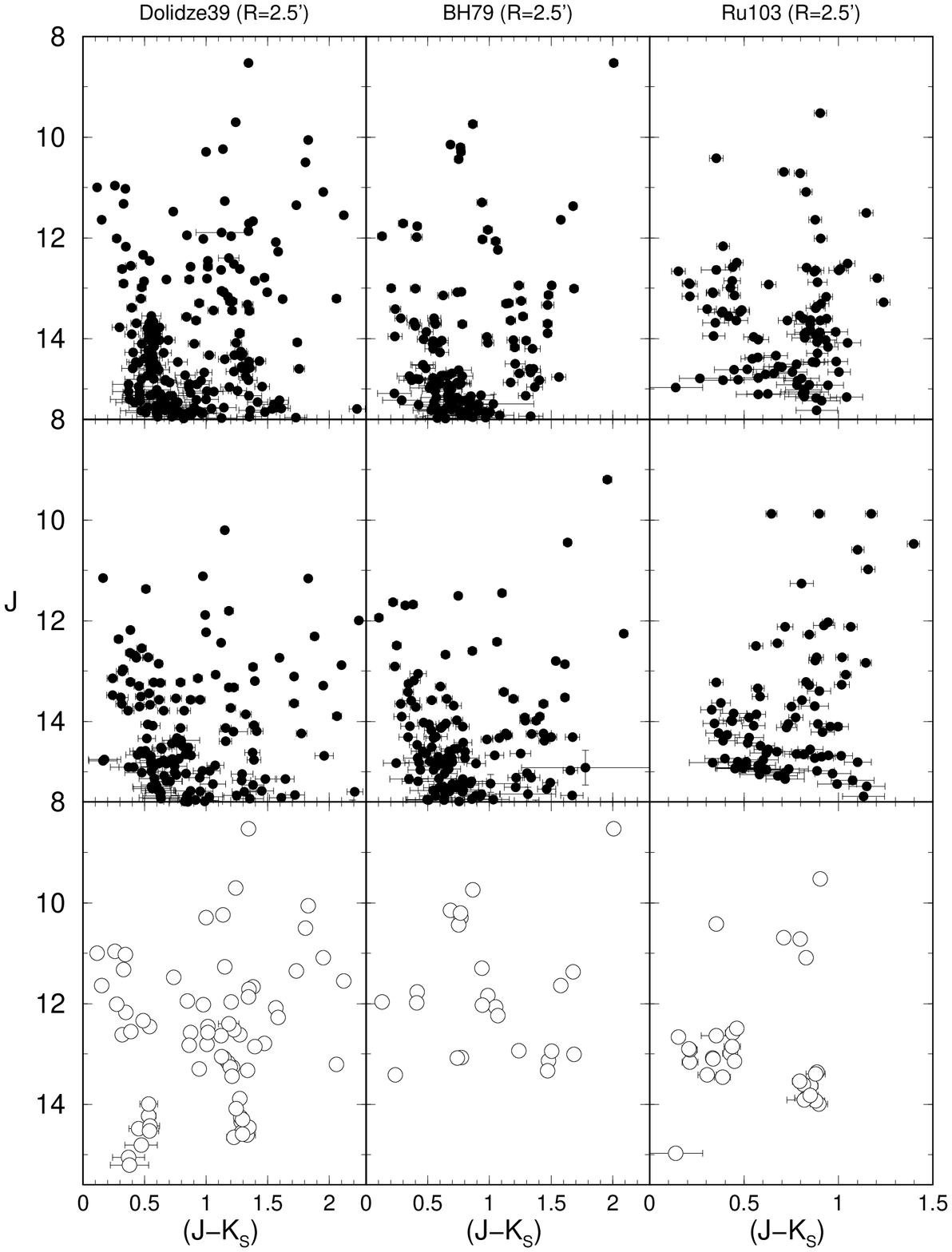}}
\caption[]{CMDs of the probable asterisms Dolidze\,39, BH\,79, and Ru\,103. }
\label{fig4}
\end{figure}

CMDs extracted from a region that contains most of the stars (see the respective RDP - 
Sect.~\ref{struc}) of each target are shown in Figs.~\ref{fig1} to \ref{fig4} top panels). 
These CMDs should be compared to those extracted from equal-area offset fields\footnote{The 
equal-area field extractions are used only for qualitative comparisons, since the 
decontamination algorithm uses a wide surrounding area (Sect.~\ref{2mass}).} (middle panels). 
Except for a few cases (e.g. Pismis\,12, IC\,1434, and Muzzio\,1) where a cluster sequence 
shows up, it is clear that the field contamination has to be taken into account for a 
proper characterisation of the target.

\subsection{Decontaminated CMDs}
\label{DecontCMD}

Our group has been developing a set of analytical tools for disentangling
cluster evolutionary sequences from field stars in CMDs. In turn, decontaminated CMDs 
have been used to investigate the nature of star cluster candidates and derive their 
astrophysical parameters. Briefly put, field-star decontamination is used to uncover 
the intrinsic CMD morphology (essential for a proper derivation of reddening, age, and 
distance from the Sun), and colour-magnitude filters are applied for building intrinsic 
stellar RDPs. In particular, the use of field-star decontamination with 2MASS photometry
in the construction of CMDs has shown to constrain age and distance significantly more 
than the observed photometry, especially for low-latitude and/or bulge-projected OCs 
(e.g. \citealt{ProbFSR}, and references therein). We apply the decontamination algorithm 
developed in \citet{BB07}\footnote{A summary of different decontamination approaches is 
given in \citet{N2244}. }. For clarity, we provide a brief description below. 

Assume that the region to decontaminate is located within $R=R_{CMD}$.
The algorithm starts by dividing the full range of magnitude and colours of a CMD 
into a 3D grid of cells with axes along the \jj\ magnitude and the \jh\ and \jk\ 
colours. Then, for each cell it computes the total (member$+$field) number-density 
of stars, $\eta_{tot}$ (for $R\la R_{CMD}$), and estimates the field-star 
number-density, $\eta_{fs}$ (for $R_{FS1}\la R\la R_{FS2}$). Next, it computes the 
expected number-density of member 
stars, $\eta_{mem}=\eta_{tot}-\eta_{fs}$, converts the number-density $\eta_{fs}$ into 
the estimated number of field stars, and subtracts it from each cell. Finally, after 
subtraction of the field stars, the remaining $N^{cell}_{clean}$ stars in each cell are 
identified for further use (see below). Photometric uncertainties (assumed to be 
Gaussian) are explicitly taken into account: we compute the probability of a star
of given magnitude and colours to be found in a any cell (i.e., the difference
of the error function computed at the cell's borders). Initially, cell dimensions 
are $\Delta\jj=1.0$ and $\Delta\jh=\Delta\jk=0.2$, but cell sizes half and twice those 
values are also used. We also apply shifts in the grid positioning by $\pm1/3$ 
of the respective cell size in the 2 colours and magnitude axes. When all the different 
setups are used, the number of independent decontamination outputs amounts to
729. 

Each setup results in a total number of member stars $N_{mem}=\sum_{cell}N^{cell}_{clean}$,
and the average of $N_{mem}$ over all setups produces the expected total number of member 
stars $\left<N_{mem}\right>$. Each star (identified above) is ranked according to the number 
of times it survives after all runs (survival frequency). Only the $\left<N_{mem}\right>$ 
highest ranked stars are considered cluster members and transposed to the respective 
decontaminated CMD. The difference between the expected number of field stars (usually 
fractional) and the actual number of stars (integer) subtracted from each cell, summed 
over all cells, is the subtraction efficiency (Further details on the decontamination 
statistics are in \citealt{BB07}). The above setup, applied to the present targets, 
resulted in  subtraction efficiencies higher than 95\%. The decontaminated CMDs are shown 
in the bottom panels of Figs.~\ref{fig1} to \ref{fig4}.

\section{Derivation of fundamental parameters}
\label{DFP}

Fundamental parameters are derived with Padova isochrones (\citealt{Girardi2002})\footnote{{\em 
http://stev.oapd.inaf.it/cgi-bin/cmd} - computed for the 2MASS filters, these isochrones are very 
similar to the Johnson-Kron-Cousins ones (e.g. \citealt{BesBret88}), with differences of at most 
0.01\,mag in colour (\citealt{TheoretIsoc}).}. The isochrones of \citet{Siess2000} are used for 
characterising the PMS stars. With respect to metallicity, the difference between, e.g. solar 
and half-solar metallicity isochrones for a given age is small, to within the 2MASS photometric 
uncertainties (\citealt{OvrlTeut}). Thus, we adopt the solar metallicity isochrones for simplicity.

The isochrone solutions take the constraints provided by the field-decontaminated CMD morphology 
into account, allowing as well for variations due to photometric uncertainties (which are small, 
because of the restrictions imposed in Sect.~\ref{2mass}), the presence of binaries (which tend 
to produce a redwards bias in the MS), and differential reddening (especially for Muzzio\,1 and
Mayer\,1). We begin with the isochrones set for zero distance modulus and reddening, and apply 
shifts in magnitude and colour until a satisfactory (judged by visual inspection) fit is obtained. 
In this sense, any isochrone that occurs within the photometric error bars is taken as acceptable, 
and the adopted age and uncertainty simply reflects the acceptable range. The adopted fits are 
shown in Figs.~\ref{fig1} to \ref{fig3}, and the respective parameters are given in Table~\ref{tab1}.

A first look at the decontaminated CMDs suggests OCs in a wide variety of evolutionary stages together
with some asterisms. In particular, the presence of somewhat evolved (in different degrees) OCs is 
suggested by the well-defined CMD morphologies containing giant clumps and red giant branches that 
show up in some clusters (Figs.~\ref{fig1} - \ref{fig3}). This is particularly true for Pismis\,12 
($\ds\sim1.9$\,kpc, $\sim1.3$\,Gyr), IC\,1434 ($\sim2.6$\,kpc, $\sim800$\,Myr), Juchert\,10 
($\sim3.2$\,kpc, $\sim800$\,Myr), and Ru\,30 ($\sim5.4$\,kpc, $\sim400$\,Myr). Pismis\,12, IC\,1434, 
Juchert\,10, and Ru\,30 also appear to contain some blue-stragglers. To this group we can also add
Muzzio\,1 ($\sim1.3$\,kpc, $\sim5$\,Myr) and Mayer\,1 ($\sim2.2$\,kpc, $\sim10$\,Myr), which are ECs 
still containing plenty PMS stars (Fig.~\ref{fig2}). Less populous but still showing evidence of 
being OCs are NGC\,3519 ($\sim2.0$\,kpc, $\sim400$\,Myr), the relatively nearby
Herschel\,1 ($\sim0.35$\,kpc, $\sim300$\,Myr) and the distant and oldest case of the sample, Ru\,140 
($\sim3.5$\,kpc, $\sim4$\,Gyr). At a lower degree we have the borderline cases, i.e. those with ambiguous
(i.e. morphology not well defined, low statistics, etc) CMDs: Ru\,146 ($\sim1.9$\,kpc, $\sim1$\,Gyr), 
Ru\,130 ($\sim0.9$\,kpc, $\sim100$\,Myr), and Ru\,129 ($\sim2.7$\,kpc, $\sim500$\,Myr). Finally, 
Dolidze\,39, BH\,79, and Ru\,103 appear to be asterisms (Fig.~\ref{fig4}). In Sect.~\ref{Conclu} 
we will couple the CMD properties to the structural analysis for establishing the nature of the targets.

\begin{figure}
\resizebox{\hsize}{!}{\includegraphics{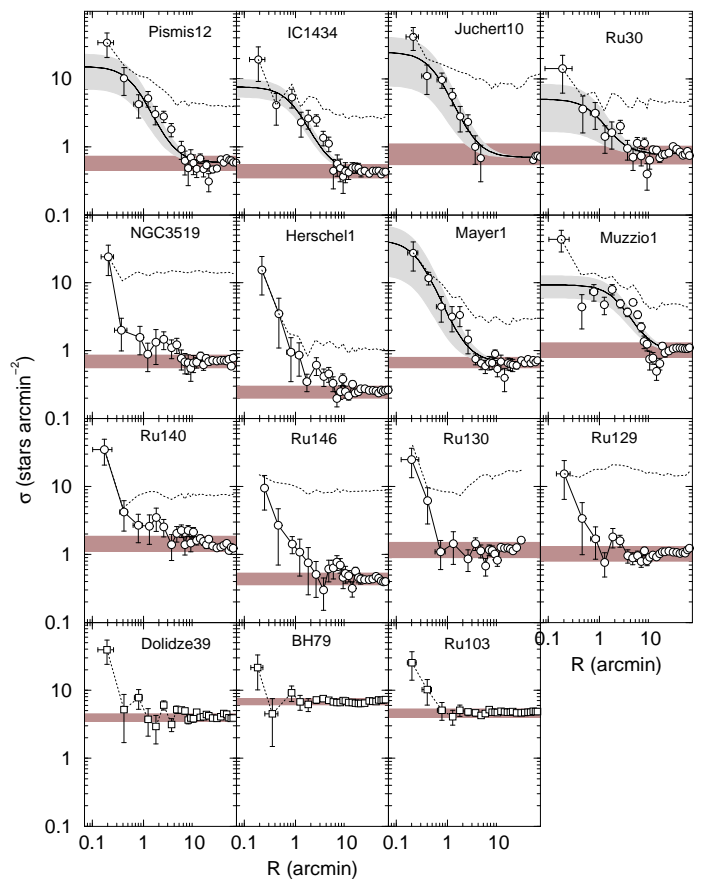}}
\caption[]{Stellar RDPs before (dotted line) and after (circles) applying the colour-magnitude
filters. Also shown is the best-fit King-like profile (solid line), the $1\sigma$ uncertainty 
(light-shaded region) and the residual background level (shaded polygon). The gap in the RDP
of Juchert\,10 is from a ``hole'' in 2MASS photometry, due to some very bright stars projected
nearby.}
\label{fig5}
\end{figure}

Muzzio\,1 and Mayer\,1 present CMDs dominated by PMS stars (Fig.~\ref{fig2}). In both cases, the rather 
poorly-populated and nearly vertical (decontaminated) MS accepts any isochrone of age within the range 
1---10\,Myr as a reasonable fit. A similar age spread is obtained for the PMS stars (allowing as well 
for the differential reddening), which are basically contained within the 0.2\,Myr and 10\,Myr (20\,Myr 
in the case of Mayer\,1) isochrones. This means that the first stars may have started to form $\sim10$\,Myr 
and $\sim20$\,Myr ago in Muzzio\,1 and Mayer\,1, respectively, with an equivalent time spread for the
star formation. Thus, we adopt $\sim5$\,Myr as the age of the bulk of the stars in Muzzio\,1, and 10\,Myr 
for Mayer\,1. 

\section{Structural analysis}
\label{struc}

Probable member stars of each target are isolated by applying the respective 
colour-magnitude filter\footnote{Which is wide enough to include MS (and PMS,
for the very young clusters) stars, photometric uncertainties, and binaries.} 
(Figs.~\ref{fig1} to \ref{fig3}). These filters are designed according to the most 
probable CMD morphology; essentially, they surround the decontaminated CMD
sequences. These stars are used to build the RDP, which is the projected stellar 
radial density profile. In previous papers (e.g. \citealt{vdB92}), we have shown that 
this procedure enhances the RDP contrast relative to the background, especially in 
crowded fields. 

Both sets of RDPs (before and after applying the colour-magnitude filters) are shown 
in Fig.~\ref{fig5}, where the contrast enhancement due to the filtering is clearly 
seen in varying degrees. We have tried to fit these RDPs with the function 
$\sigma(R)=\sigma_{bg}+\sigma_0/(1+(R/R_c)^2)$, where $\sigma_0$ and $\sigma_{bg}$ are 
the central and residual fore/background stellar densities, and \rc\ is the core radius. 
This function is similar to that applied by \cite{King1962} to the surface-brightness 
profiles in the central parts of globular clusters, although here we use it on star 
counts\footnote{Given the relatively low number of member stars, RDPs are expected to 
present lower fluctuations than surface-brightness profiles (\citealt{StrucPar}).}. 
However, only in 6 cases (IC\,1434, Pismis\,12, Juchert\,10, Ru\,30, Mayer\,1, and 
Muzzio\,1) the fitting converged towards a meaningful solution. The best-fit solutions 
are shown in Fig.~\ref{fig5}, and the structural parameters are given in Table~\ref{tab2}. 
NGC\,3519, Herschel\,1, and Ru\,146, have RDPs that marginally could be represented by 
the King-like function, but the error bars and RDP fluctuations prevented that. To a 
lesser degree, the same applies to Ru\,129,130,140. On the other hand, Dolidze\,39, BH\,79, 
and Ru\,103, have RDPs typical of asterisms (e.g. \citealt{FSR20}).

\begin{table*}
\caption[]{Structural parameters derived from the RDPs}
\label{tab2}
\renewcommand{\tabcolsep}{3.1mm}
\renewcommand{\arraystretch}{1.0}
\begin{tabular}{cccccccccc}
\hline\hline
Cluster&$\sigma_0$&\rc&\rl&$1\arcmin$&$\sigma_0$&\rc&\rl&$N^*_C$\\
       &$\rm(*\,\arcmin^{-2})$&(\arcmin)&(\arcmin) &(pc)&$\rm(*\,pc^{-2})$&(pc)&(pc)&(stars)\\
(1)&(2)&(3)&(4)&(5)&(6)&(7)&(8)&(9)\\
\hline
\multicolumn{9}{c}{Open clusters}\\
\hline
IC\,1434 &$7.2\pm2.2$&$0.90\pm0.23$&$5.0\pm0.5$&0.754&$12.6\pm3.9$&$0.68\pm0.23$&$3.8\pm1.0$&$19\pm10$\\
Mayer\,1 &$41.4\pm28.9$&$0.27\pm0.12$&$3.5\pm0.5$&0.645&$99.6\pm69.6$&$0.17\pm0.10$&$2.3\pm0.9$&$10\pm9$\\
Herschel\,1 &---&---&$10.0\pm2.0$&0.100&---&---&$1.0\pm0.3$&---\\
Ruprecht\,30 &$4.3\pm3.3$&$0.83\pm0.52$&$3.5\pm0.5$&1.571&$1.8\pm1.3$&$1.31\pm0.92$&$5.5\pm1.5$&$9\pm13$\\
Muzzio\,1 &$8.3\pm3.3$&$2.31\pm0.74$&$8.5\pm0.5$&0.385&$55.9\pm22.2$&$0.90\pm0.35$&$3.3\pm0.8$&$140\pm103$\\
Pismis\,12 &$14.6\pm8.0$&$0.66\pm0.37$&$5.5\pm0.5$&0.549&$48.4\pm26.5$&$0.36\pm0.21$&$3.0\pm0.6$&$20\pm24$\\
NGC\,3519 &---&---&$5.0\pm0.5$&0.585&---&---&$2.9\pm0.8$&---\\
Juchert\,10 &$23.9\pm16.6$&$0.55\pm0.27$&$3.0\pm0.5$&0.931&$27.6\pm19.1$&$0.51\pm0.28$&$2.8\pm0.8$&$23\pm26$\\
\hline
\multicolumn{9}{c}{Borderline cases}\\
\hline
Ruprecht\,140 &---&---&$3.0\pm0.5$&1.006&---&---&$3.0\pm0.8$&---\\
Ruprecht\,146 &---&---&$2.5\pm0.3$&0.546&---&---&$1.3\pm0.3$&---\\
Ruprecht\,130 &---&---&$1.3\pm0.3$&0.248&---&---&$0.3\pm0.1$&---\\
Ruprecht\,129 &---&---&$3.5\pm0.5$&0.769&---&---&$2.7\pm0.7$&---\\
\hline
\multicolumn{9}{c}{Probable asterisms}\\
\hline
Dolidze\,39 &---&---&$8.0\pm2.0$&---&---&---&---&---\\
BH\,79 &---&---&$1.5\pm0.3$&---&---&---&---&---\\
Ruprecht\,103 &---&---&$1.0\pm0.3$&---&---&---&---&---\\
\hline
\end{tabular}
\begin{list}{Table Notes.}
\item Col.~5: arcmin to parsec scale. Col.~9: Number of stars in the core 
($N^*_C=\sigma_0\times\pi\,\rc^2$) estimated from the King fit parameters. 
For comparison with other clusters, the King-like central stellar density 
($\sigma_0$), core radius (\rc), and cluster radius (\rl) are given both in 
angular and absolute units. Uncertainties in cols.~7 and 8 include the
errors in distance determination.
\end{list}
\end{table*} 

We also use the RDP and background level to estimate the cluster radius \rl\ 
(Table~\ref{tab2}), defined as the distance from the cluster centre where RDP and 
residual background are statistically indistinguishable. In this sense, \rl\ is an 
observational truncation radius, whose value depends both on the radial distribution 
of member stars and the field density. 







\section{Summary and conclusions}
\label{Conclu}

In this paper we investigate the nature of 15 challenging stellar overdensities that,
in general, have conflicting classifications (star cluster or asterism) in different 
studies. With a few exceptions, the targets are poorly-populated and/or projected against 
very dense backgrounds, which might explain the difficulty in establishing their nature. 
They are: IC\,1434, Mayer\,1, Herschel\,1, Muzzio\,1, Pismis\,12, NGC\,3519, Juchert\,10, 
Dolidze\,39, BH\,79, Ruprecht\,30, 103, 129, 130, 140, and 146.

Our approach involves photometric properties derived from field-star decontaminated CMDs, 
and structural parameters, derived from highly-contrasted and large-scale RDPs. The field 
decontamination, in particular, has proven to be essential for deeper analyses on the nature 
of the targets. This point is particularly obvious when one considers the RDPs (Fig.~\ref{fig5}).
Prior to field decontamination, most RDPs are similar to those produced by asterisms. In 
contrast, by eliminating most of the field stars, the decontaminated RDPs of a few cases 
could be represented by a King-like profile, thus suggesting a cluster. At this point
we note that an objective criterion for unambiguous classifying the overdensities would be 
interesting. However, given the rather different distances, number of member stars, field 
contamination level, etc, such a criterion based only on the RDP does not seem feasible, 
since a faint star cluster may have an RDP resembling that of an asterism. In this sense, 
we apply a subjective selection criterion based on both properties, the decontaminated CMD 
morphology and RDP. As usual, preference is given to the CMD. Based on this criterion, the 
targets can be grouped in the three categories below.

\paragraph{\bf Open and embedded clusters:} This group is composed by Pismis\,12, IC\,1434, 
Juchert\,10, Ruprecht\,30, NGC\,3519, Herschel\,1, Mayer\,1, and Muzzio\,1, the latter two 
being ECs. Their decontaminated CMDs (Figs.~\ref{fig1} and \ref{fig2}) present evolutionary 
sequences of OCs and ECs at different evolutionary stages. Their ages range from 300\,Myr to 
1.3\,Gyr, except for the young ones Mayer\,1 and Muzzio\,1 that, with ages within 5-10\,Myr, 
have CMDs dominated by PMS stars. The RDPs are well represented by a King-like profile, except 
for Herschel\,1 and NGC\,3519, whose RDPs present significant deviations. 

\paragraph{\bf Borderline cases:} Members of this group are Ruprecht\,129, 130, 140, and 146.
They have ambiguous CMDs, i.e., although rather poorly-populated, the decontaminated sequences 
appear to suggest star clusters. However, because of the low statistics, no definitive conclusion 
can be drawn. Also, their RDPs cannot be represented by the King-like profile. Thus, deeper
photometry is required.

\paragraph{\bf Probable asterisms:} Both the CMDs and RDPs of Dolidze\,39, BH\,79, and 
Ruprecht\,103, appear to arise from field fluctuations. Thus, the photometric and
structural criteria indicate that they are asterisms.

A general conclusion is that, in some cases, only a thin line separates properties of star 
clusters and asterisms (field fluctuations), owing to a low-stellar population, dense 
field-contamination residuals, and/or significant differential reddening. It is clear that 
specific analytical tools may be necessary to establish the nature of these difficult cases. 
Field-decontaminated CMDs and RDPs, which provide a deep and more constrained investigation, 
appear to be essential for this task.

\section*{Acknowledgements}
We thank an anonymous referee for important comments and suggestions.
We acknowledge support from the Brazilian Institution CNPq.
This publication makes use of data products from the Two Micron All Sky Survey, which
is a joint project of the University of Massachusetts and the Infrared Processing and
Analysis Centre/California Institute of Technology, funded by the National Aeronautics
and Space Administration and the National Science Foundation. We also employed the WEBDA 
database, operated at the Institute for Astronomy of the University of Vienna. The SIMBAD 
database, operated at CDS, Strasbourg, France, has also been useful.

\appendix
\section{DSS/XDSS images}
\label{appA}

The images have been taken from CADC Sky Survey, with a field of view adequate
to the angular dimension of each target. The same applies to the image
band. 



\begin{figure*}
\begin{minipage}[b]{0.33\linewidth}
\includegraphics[width=\textwidth]{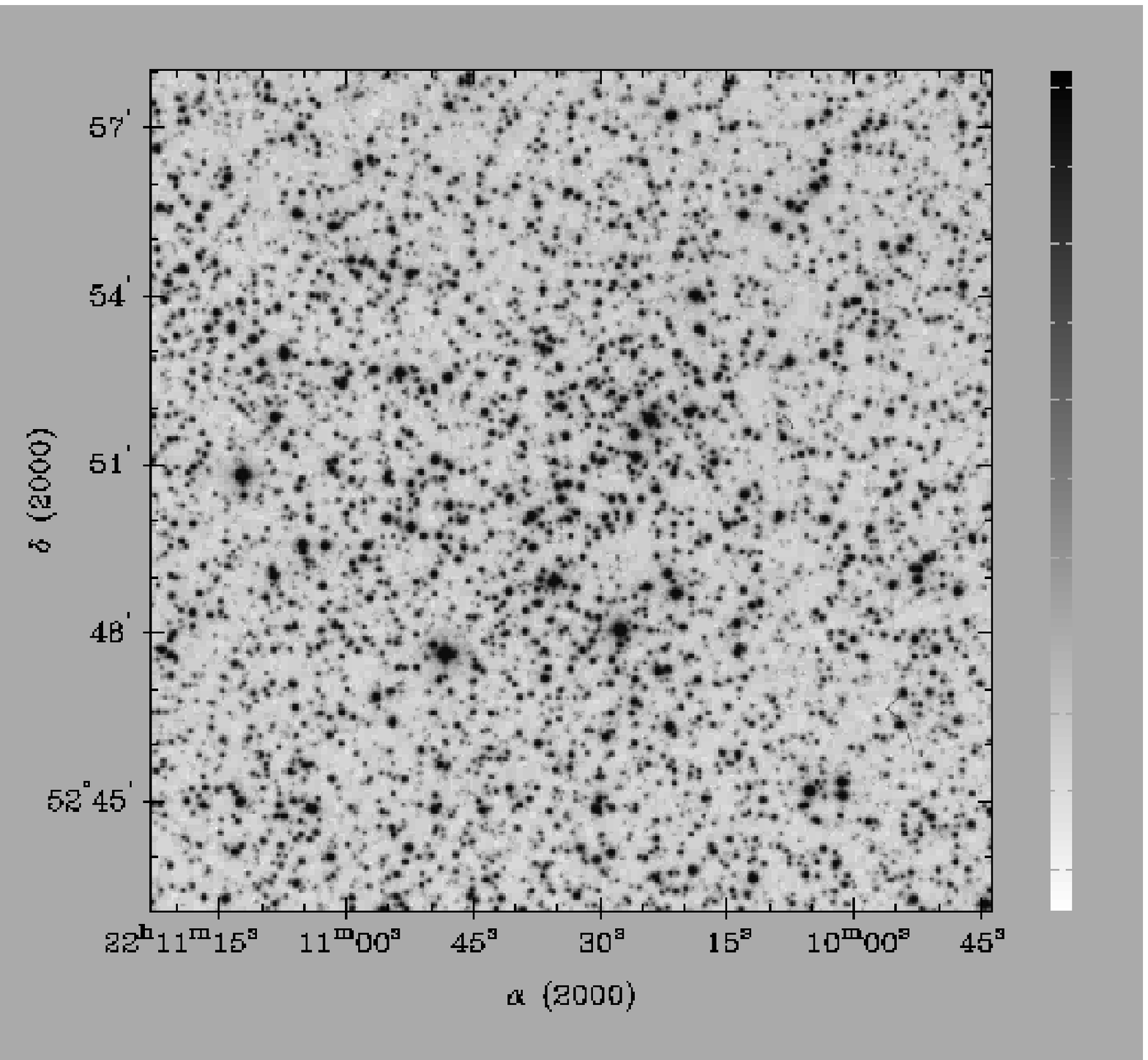}
\end{minipage}\hfill
\begin{minipage}[b]{0.33\linewidth}
\includegraphics[width=\textwidth]{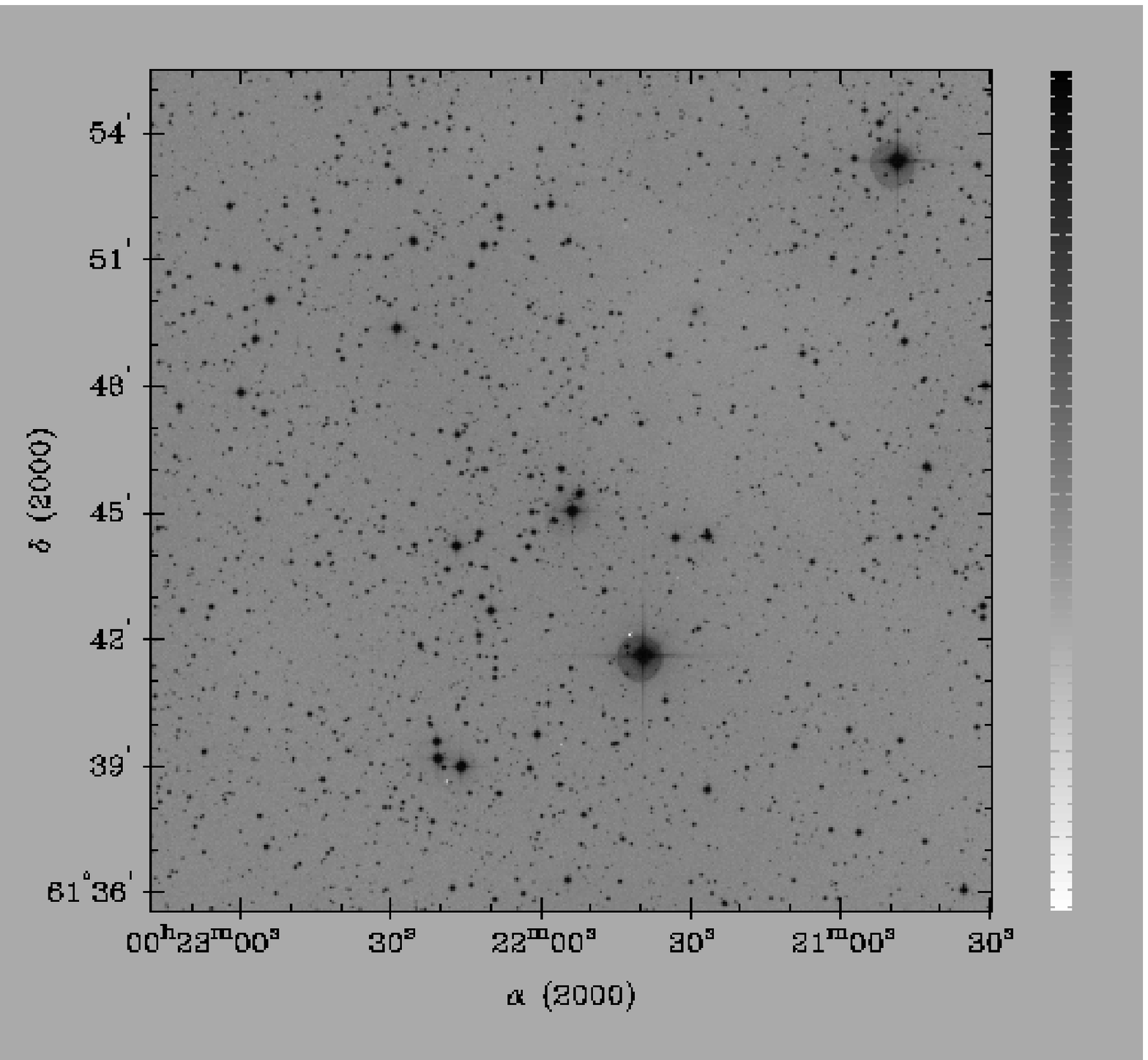}
\end{minipage}\hfill
\begin{minipage}[b]{0.33\linewidth}
\includegraphics[width=\textwidth]{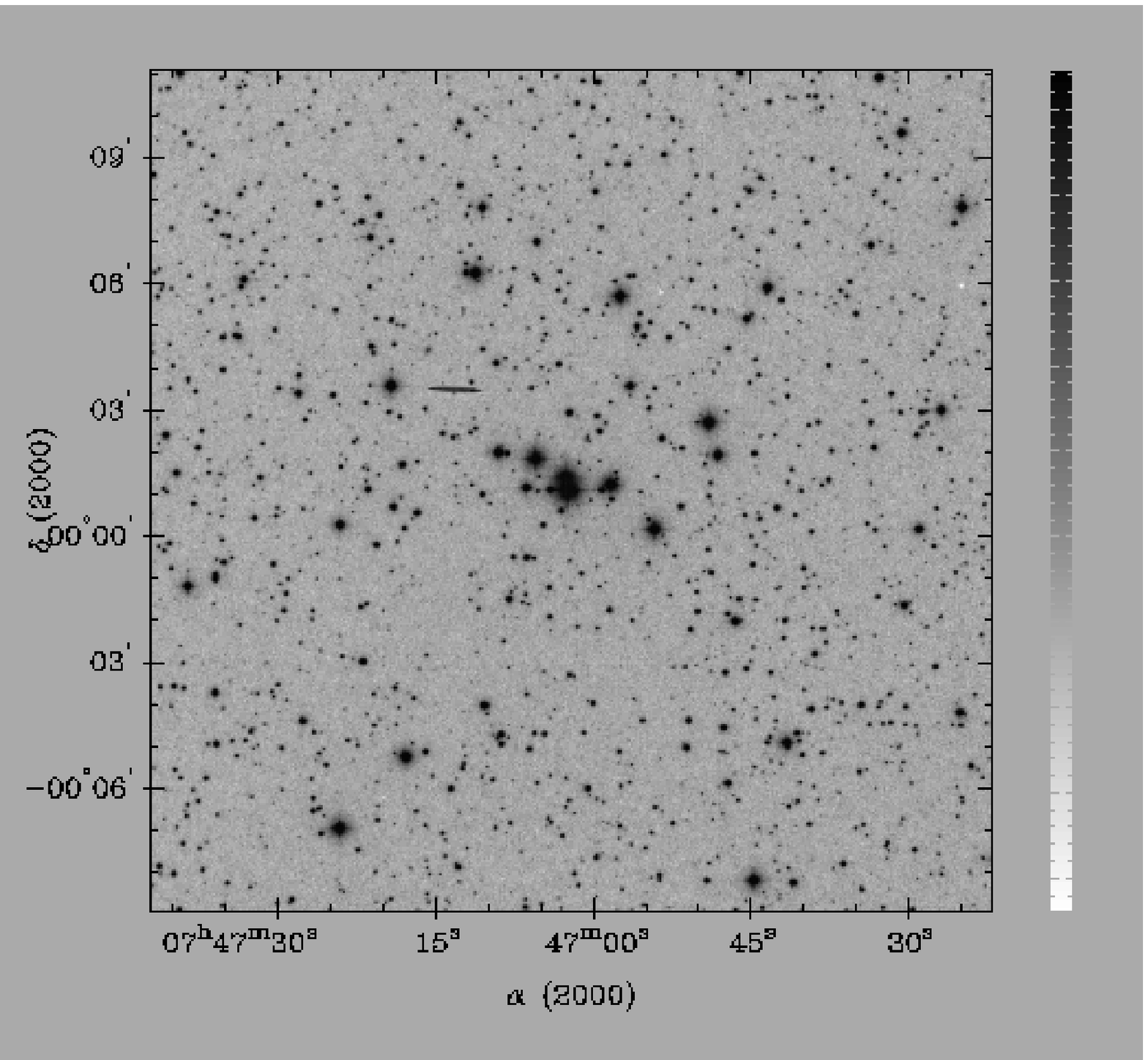}
\end{minipage}\hfill

\begin{minipage}[b]{0.33\linewidth}
\includegraphics[width=\textwidth]{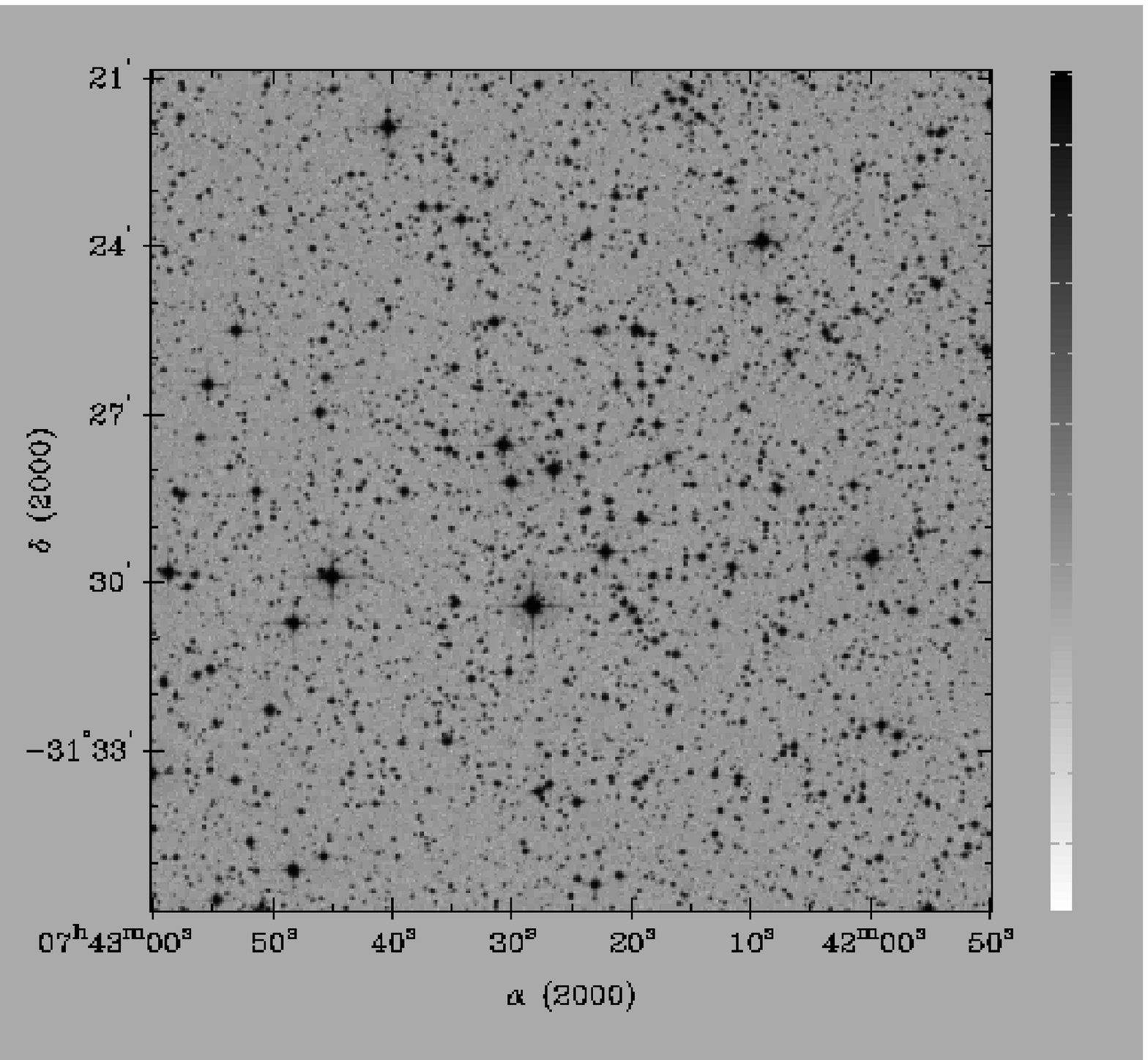}
\end{minipage}\hfill
\begin{minipage}[b]{0.33\linewidth}
\includegraphics[width=\textwidth]{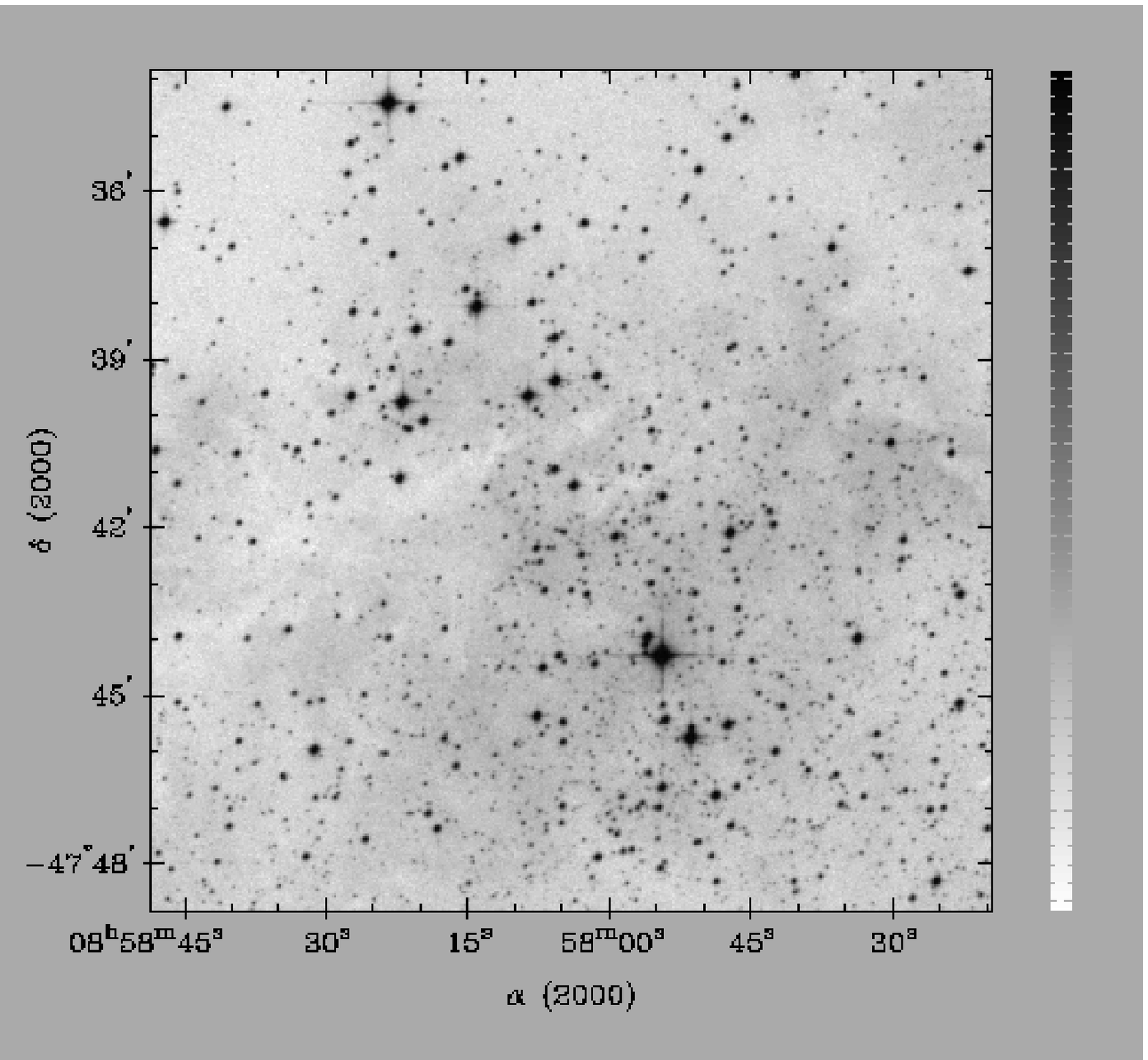}
\end{minipage}\hfill
\begin{minipage}[b]{0.33\linewidth}
\includegraphics[width=\textwidth]{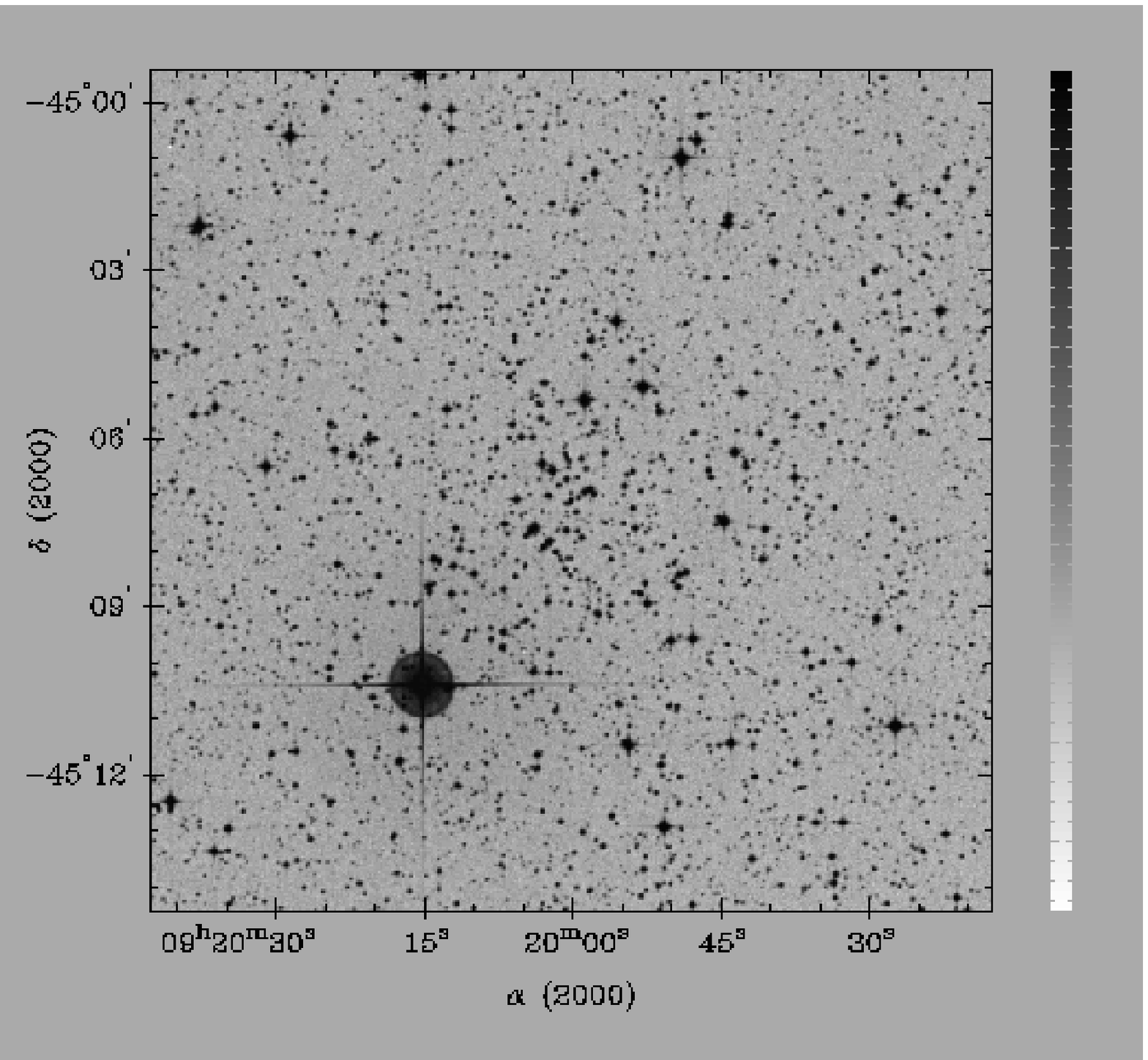}
\end{minipage}\hfill

\begin{minipage}[b]{0.33\linewidth}
\includegraphics[width=\textwidth]{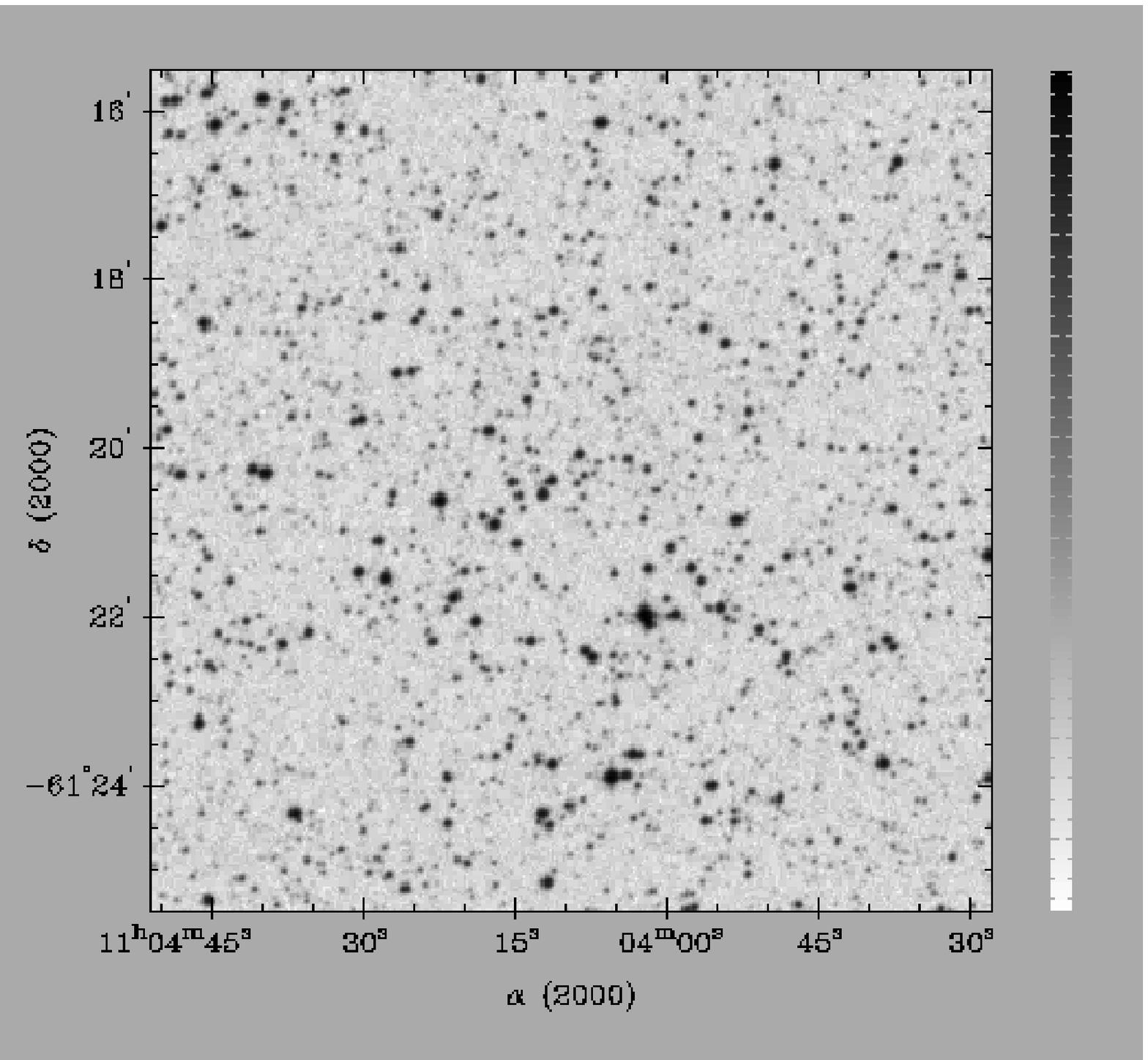}
\end{minipage}\hfill
\begin{minipage}[b]{0.33\linewidth}
\includegraphics[width=\textwidth]{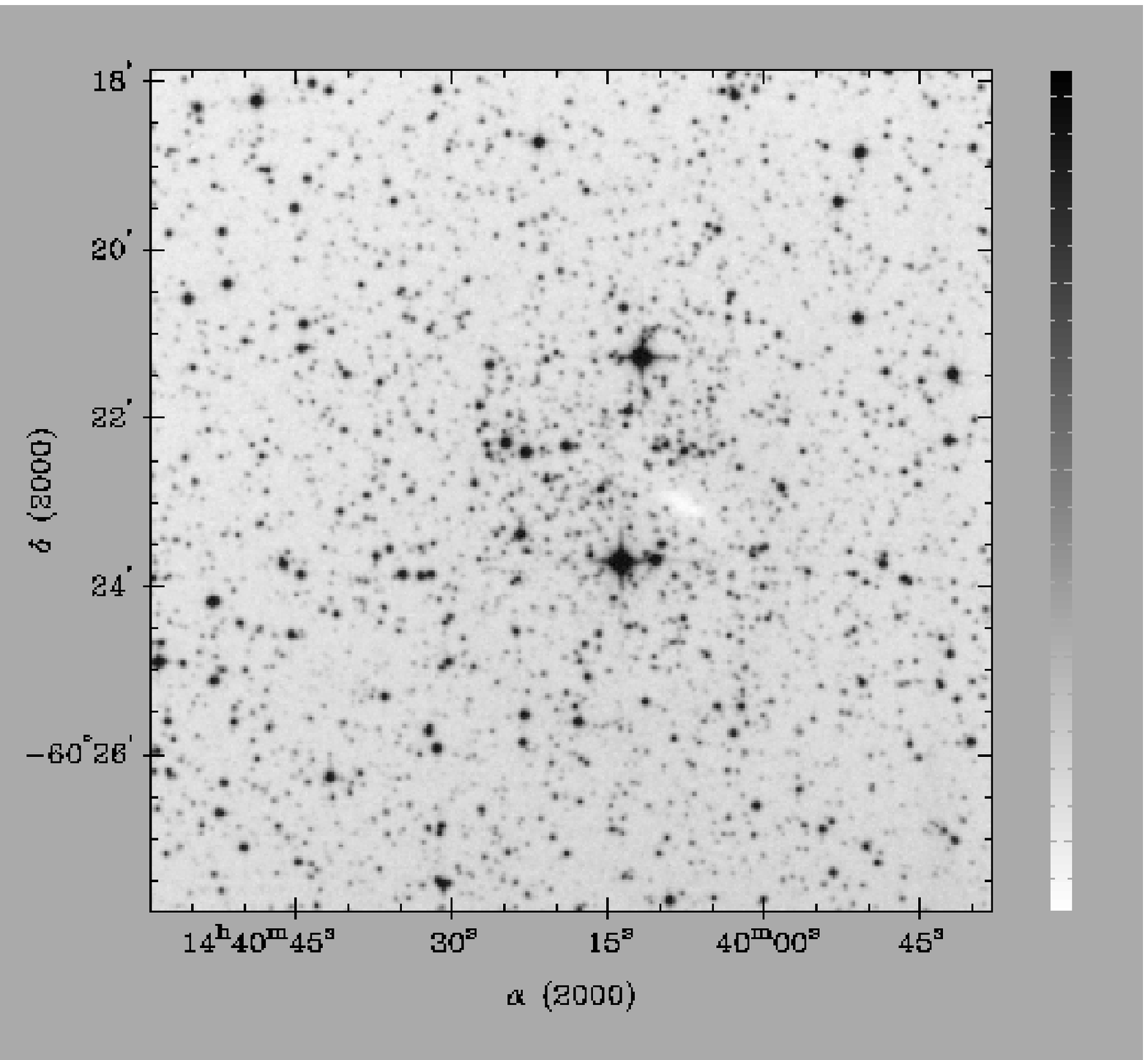}
\end{minipage}\hfill
\begin{minipage}[b]{0.33\linewidth}
\includegraphics[width=\textwidth]{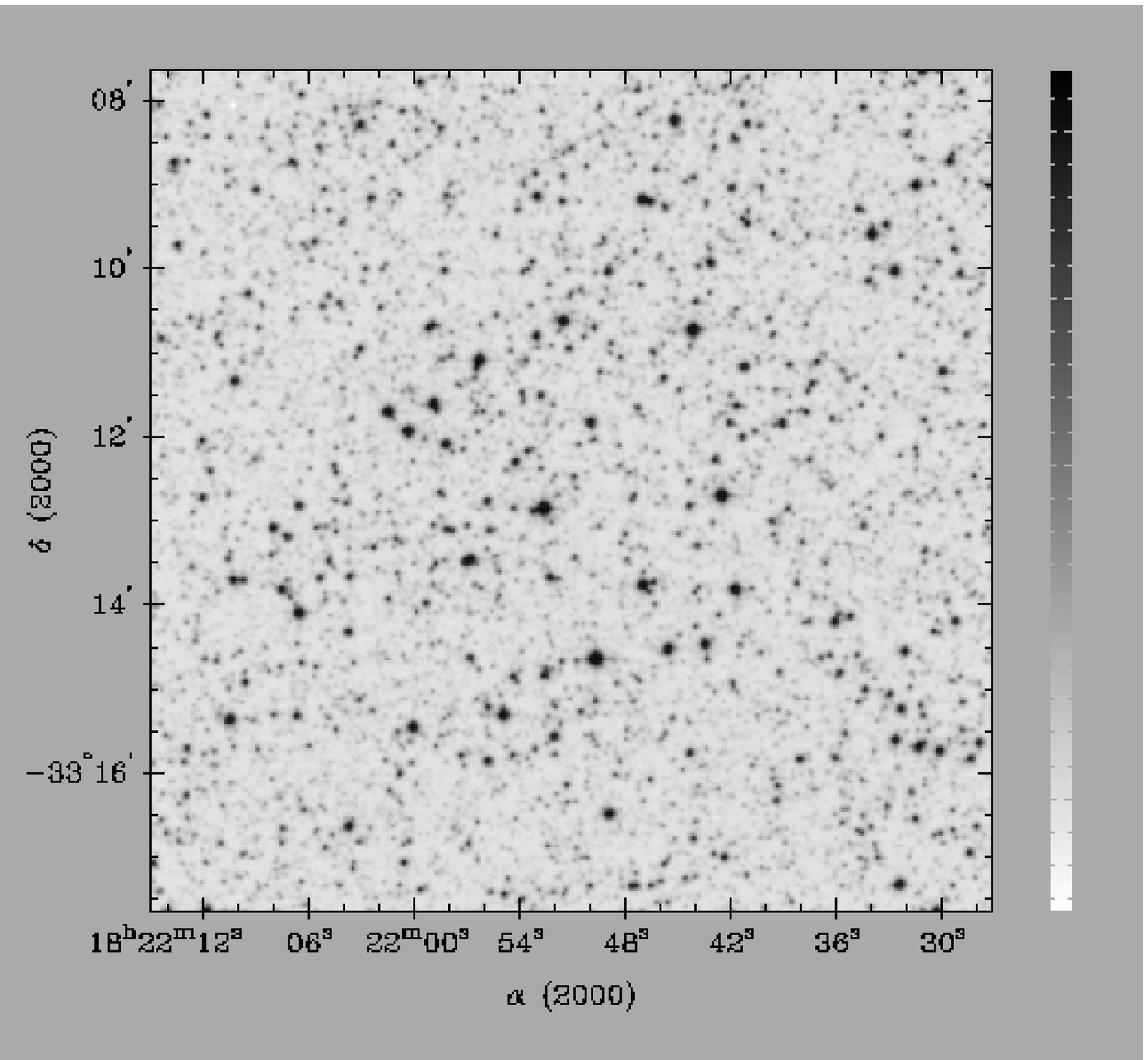}
\end{minipage}\hfill

\begin{minipage}[b]{0.33\linewidth}
\includegraphics[width=\textwidth]{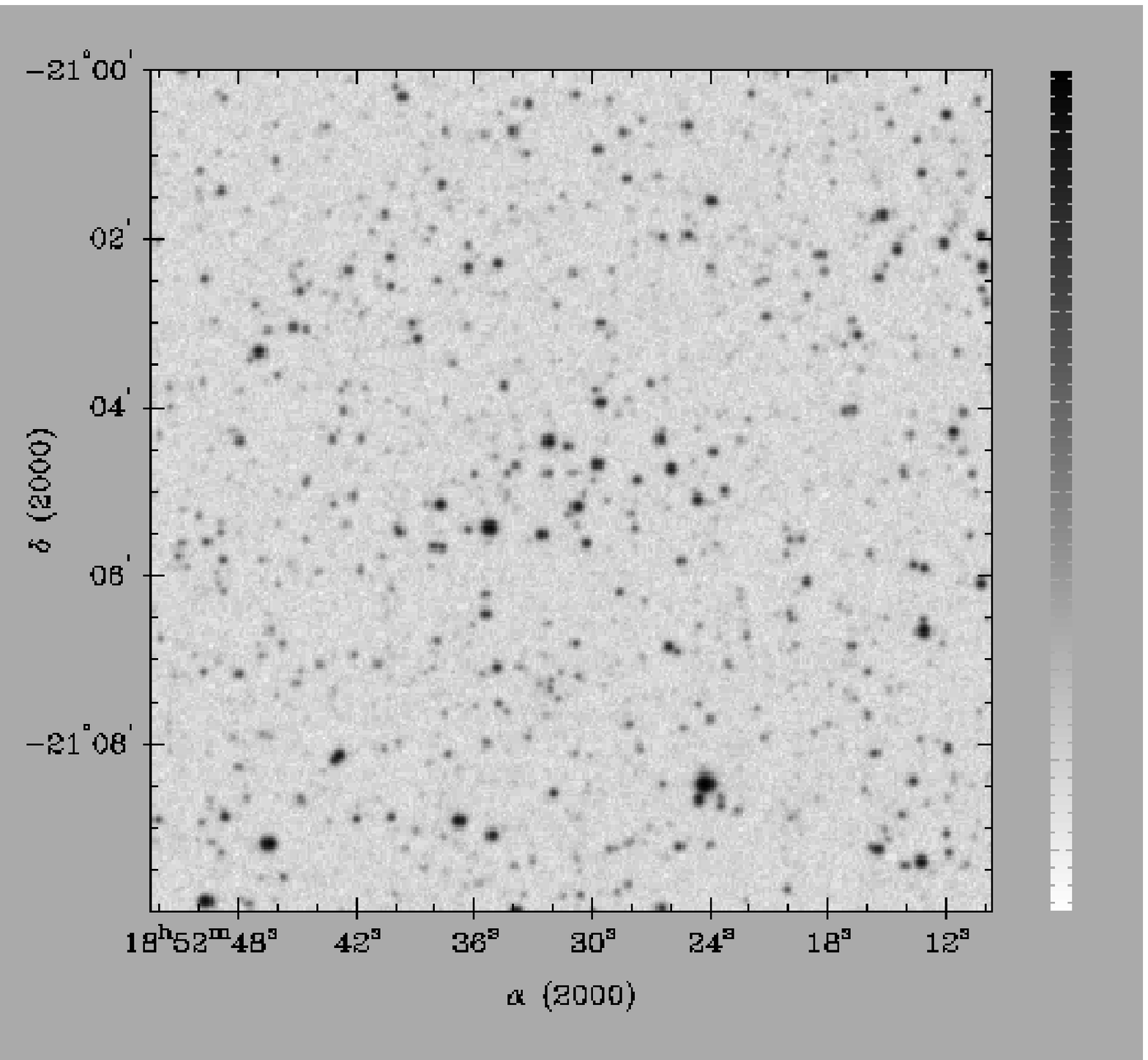}
\end{minipage}\hfill
\begin{minipage}[b]{0.33\linewidth}
\includegraphics[width=\textwidth]{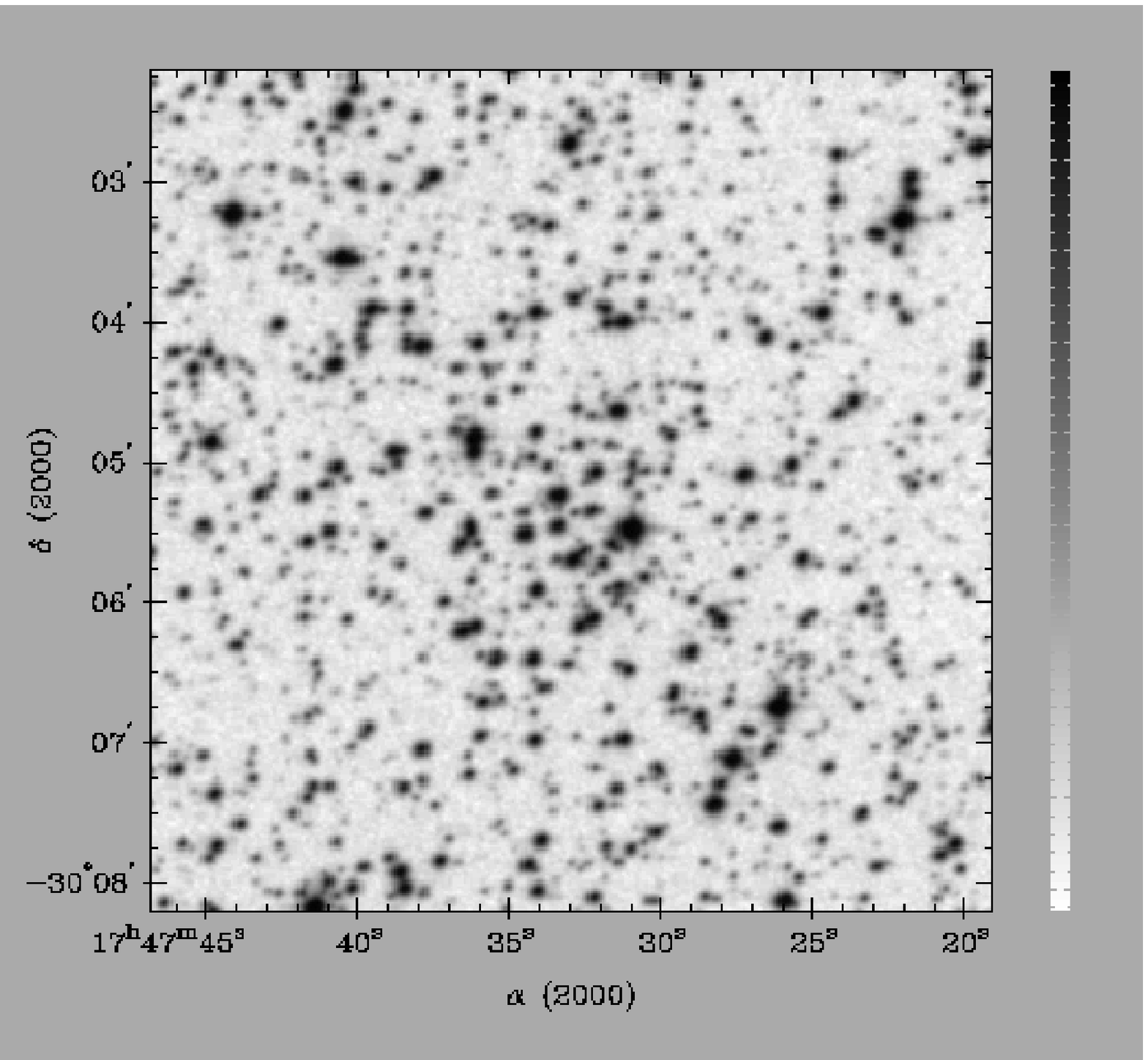}
\end{minipage}\hfill
\begin{minipage}[b]{0.33\linewidth}
\includegraphics[width=\textwidth]{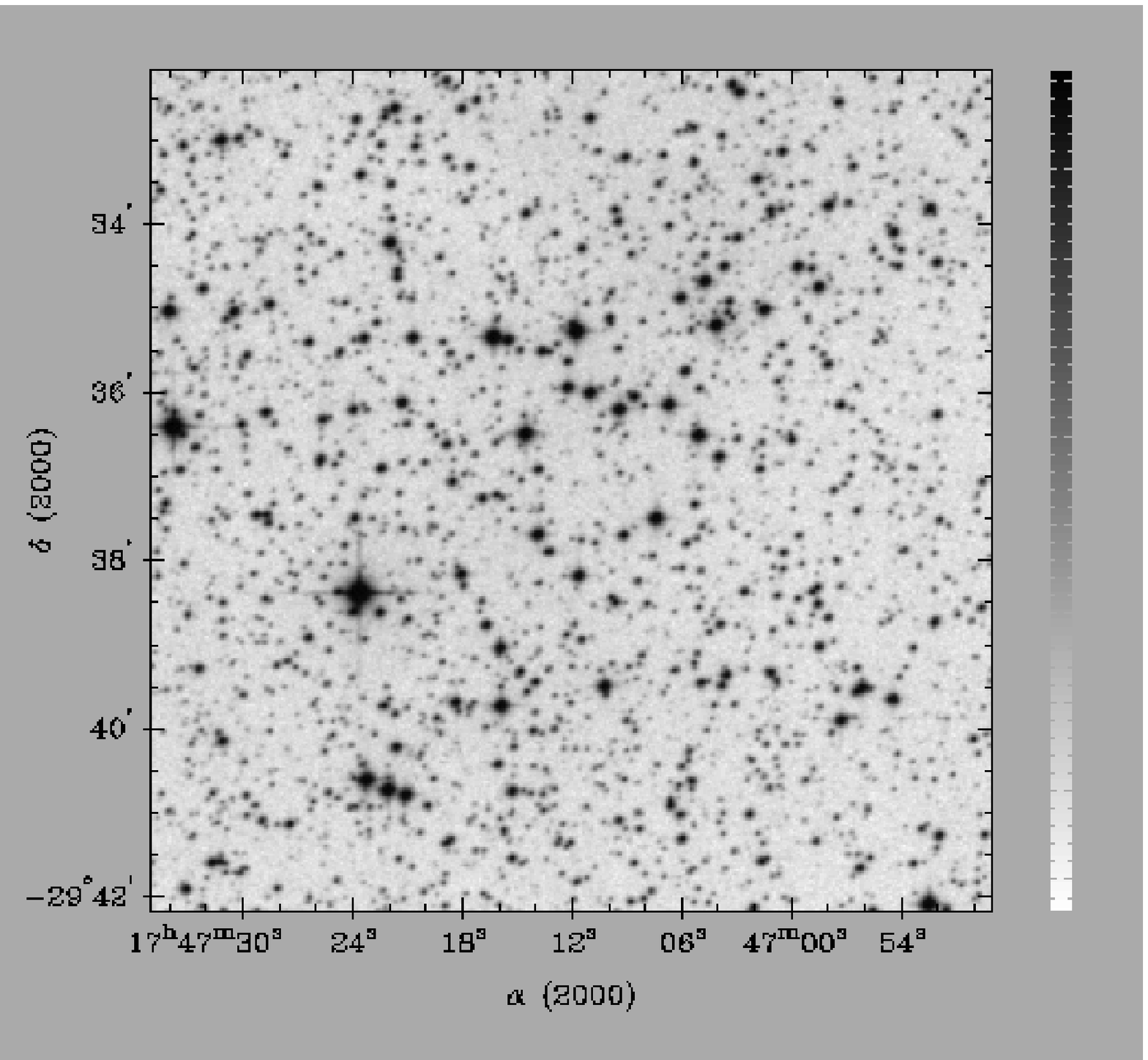}
\end{minipage}\hfill
\caption[]{CADC images of the targets. Clusters are shown from left to right.
First row: IC\,1434 (B, $15\arcmin\times15\arcmin$), Mayer\,1 (B, $20\arcmin\times20\arcmin$), Herschel\,1 (B, $20\arcmin\times20\arcmin$); 
Second row: Ru\,30 (B, $10\arcmin\times10\arcmin$), Muzzio\,1 (R, $15\arcmin\times15\arcmin$), Pismis\,12 (B, $10\arcmin\times10\arcmin$); 
Third row: NGC\,3519 (B, $10\arcmin\times10\arcmin$), Juchert\,10 (I, $10\arcmin\times10\arcmin$), Ru\,140 (I, $10\arcmin\times10\arcmin$); 
Fourth row: Ru\,146 (B, $10\arcmin\times10\arcmin$), Ru\,130 (R, $6\arcmin\times6\arcmin$), and Ru\,129 (R, $10\arcmin\times10\arcmin$).}
\label{fig_APA1}
\end{figure*}

\begin{figure*}

\begin{minipage}[b]{0.33\linewidth}
\includegraphics[width=\textwidth]{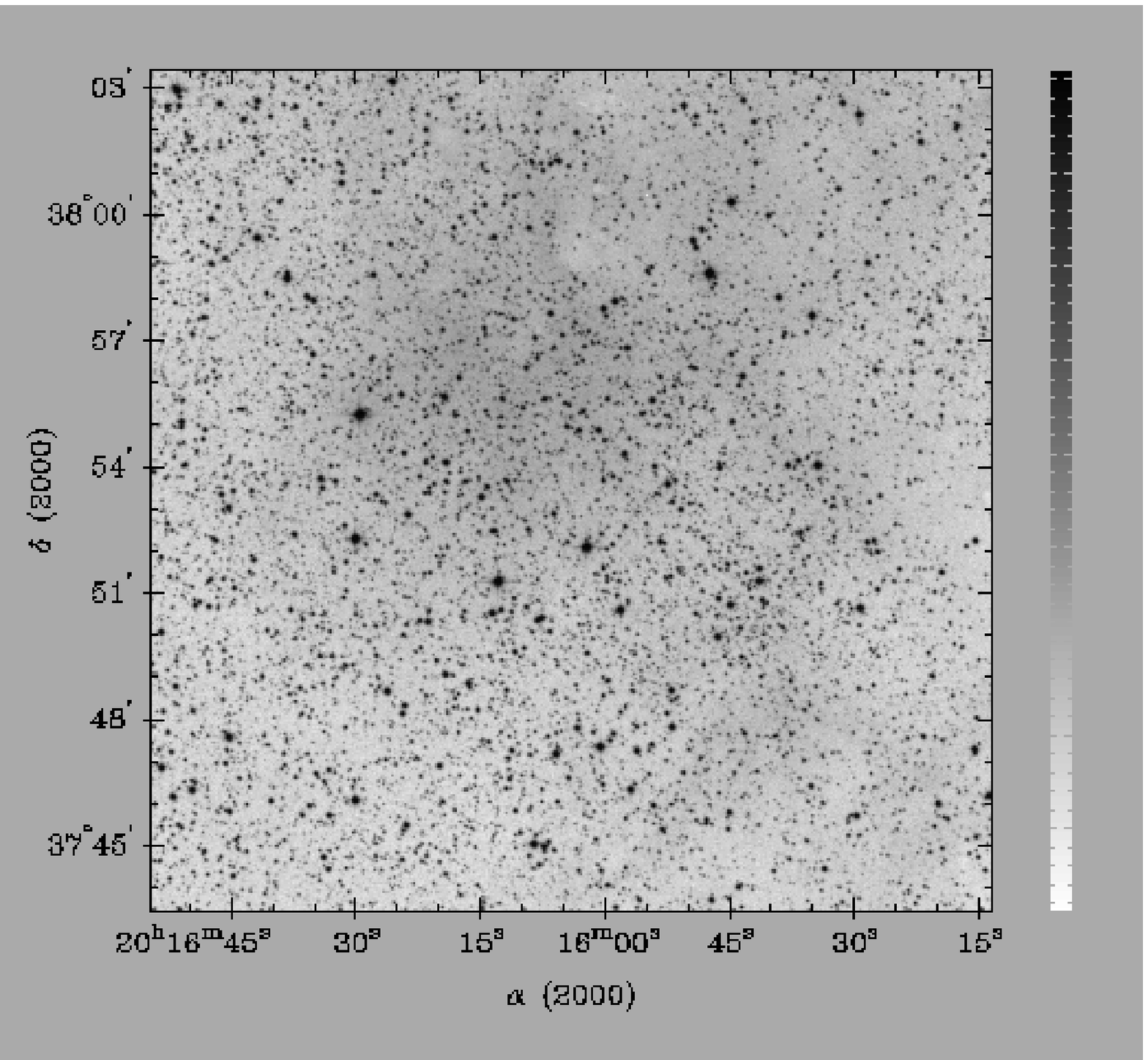}
\end{minipage}\hfill
\begin{minipage}[b]{0.33\linewidth}
\includegraphics[width=\textwidth]{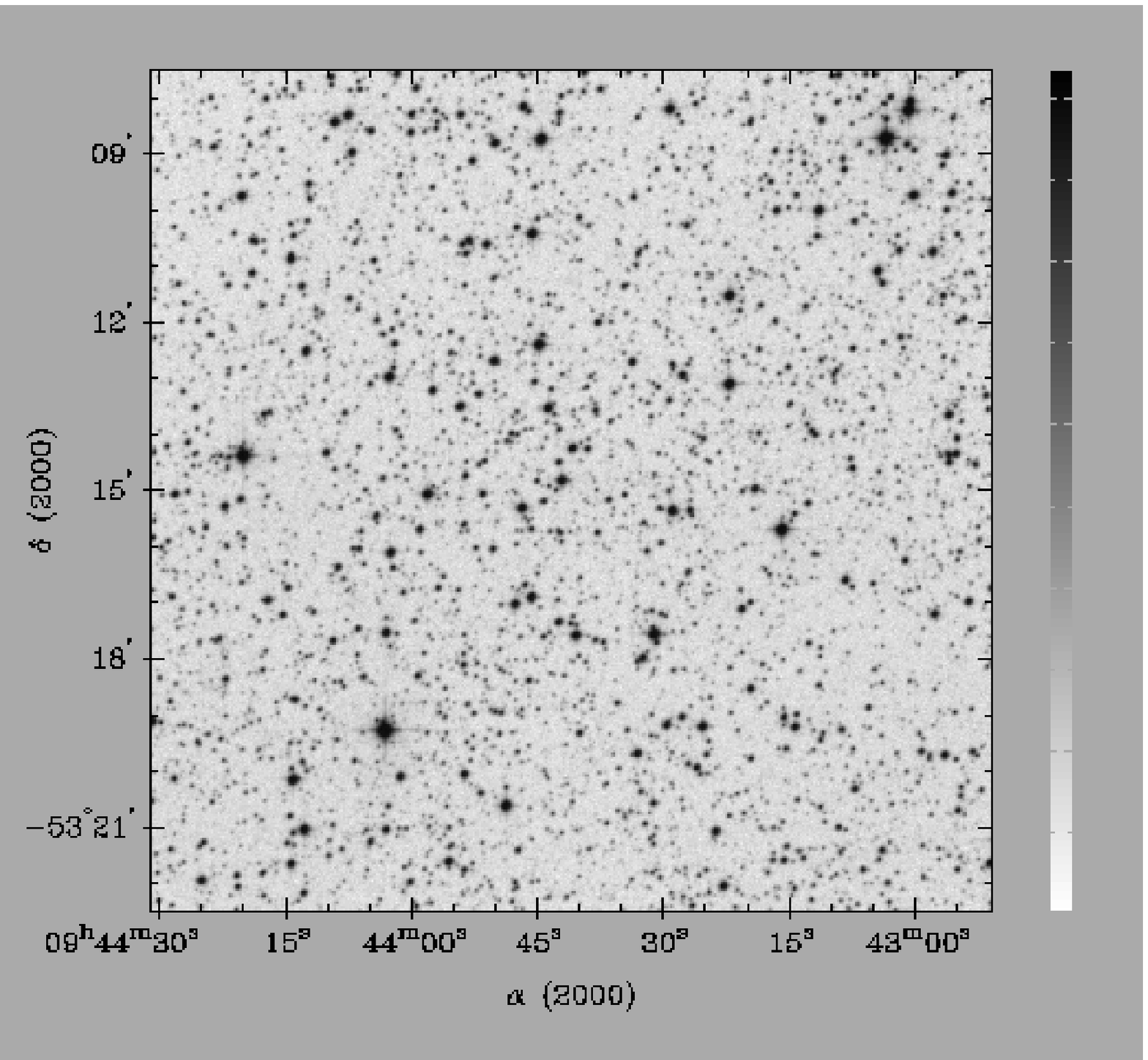}
\end{minipage}\hfill
\begin{minipage}[b]{0.33\linewidth}
\includegraphics[width=\textwidth]{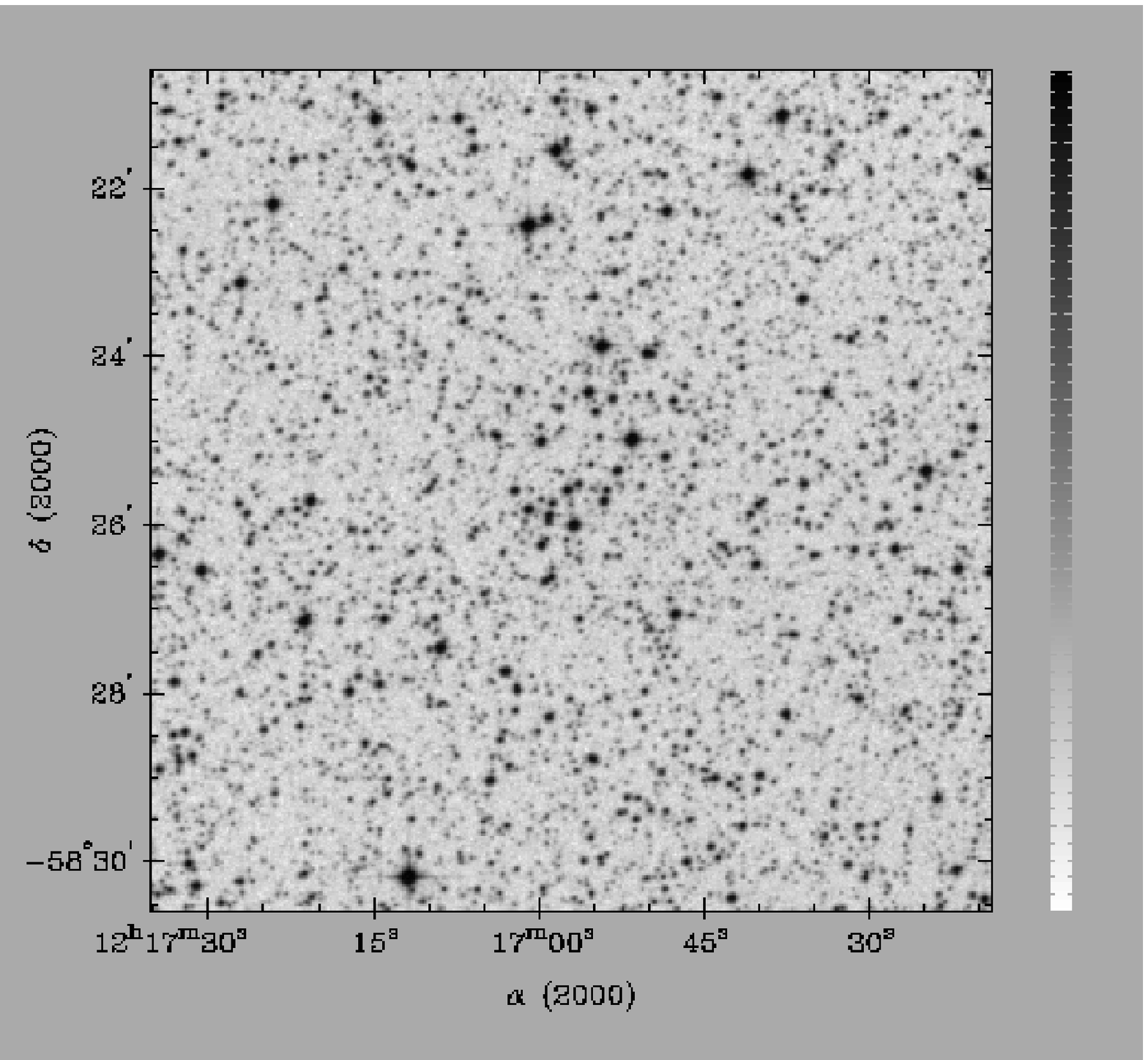}
\end{minipage}\hfill
\caption[]{Same as Fig.~\ref{fig_APA1} for Dolidze\,39 (R, $20\arcmin\times20\arcmin$),
BH\,79 (R, $15\arcmin\times15\arcmin$), and Ru\,103 (R, $10\arcmin\times10\arcmin$).}
\label{fig_APA2}
\end{figure*}

\section{2MASS images}
\label{appB}

Near-infrared (\ks) images of optically absorbed clusters (Table~\ref{tab1}). 

\begin{figure*}
\begin{minipage}[b]{0.33\linewidth}
\includegraphics[width=\textwidth]{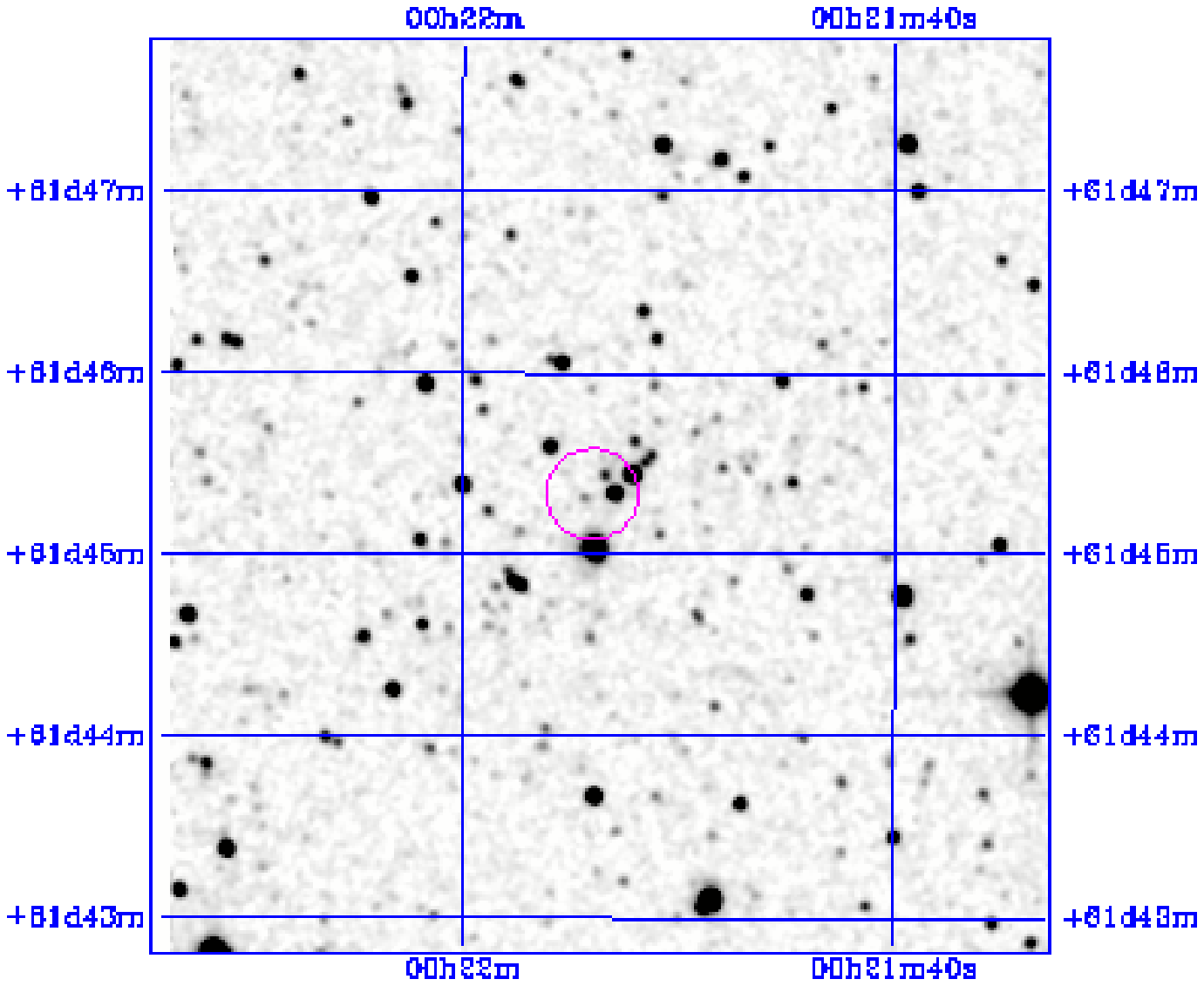}
\end{minipage}\hfill
\begin{minipage}[b]{0.33\linewidth}
\includegraphics[width=\textwidth]{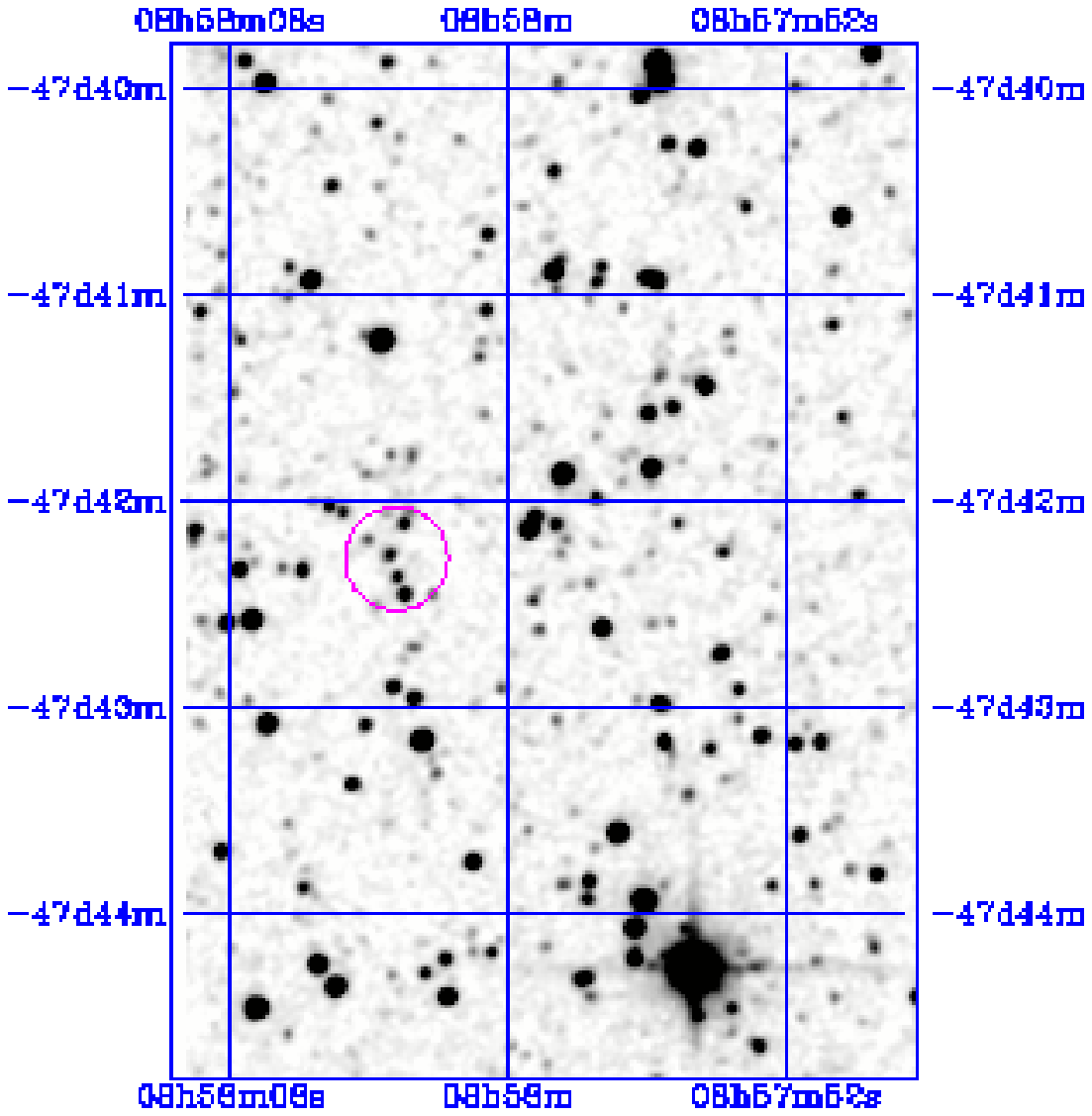}
\end{minipage}\hfill
\begin{minipage}[b]{0.33\linewidth}
\includegraphics[width=\textwidth]{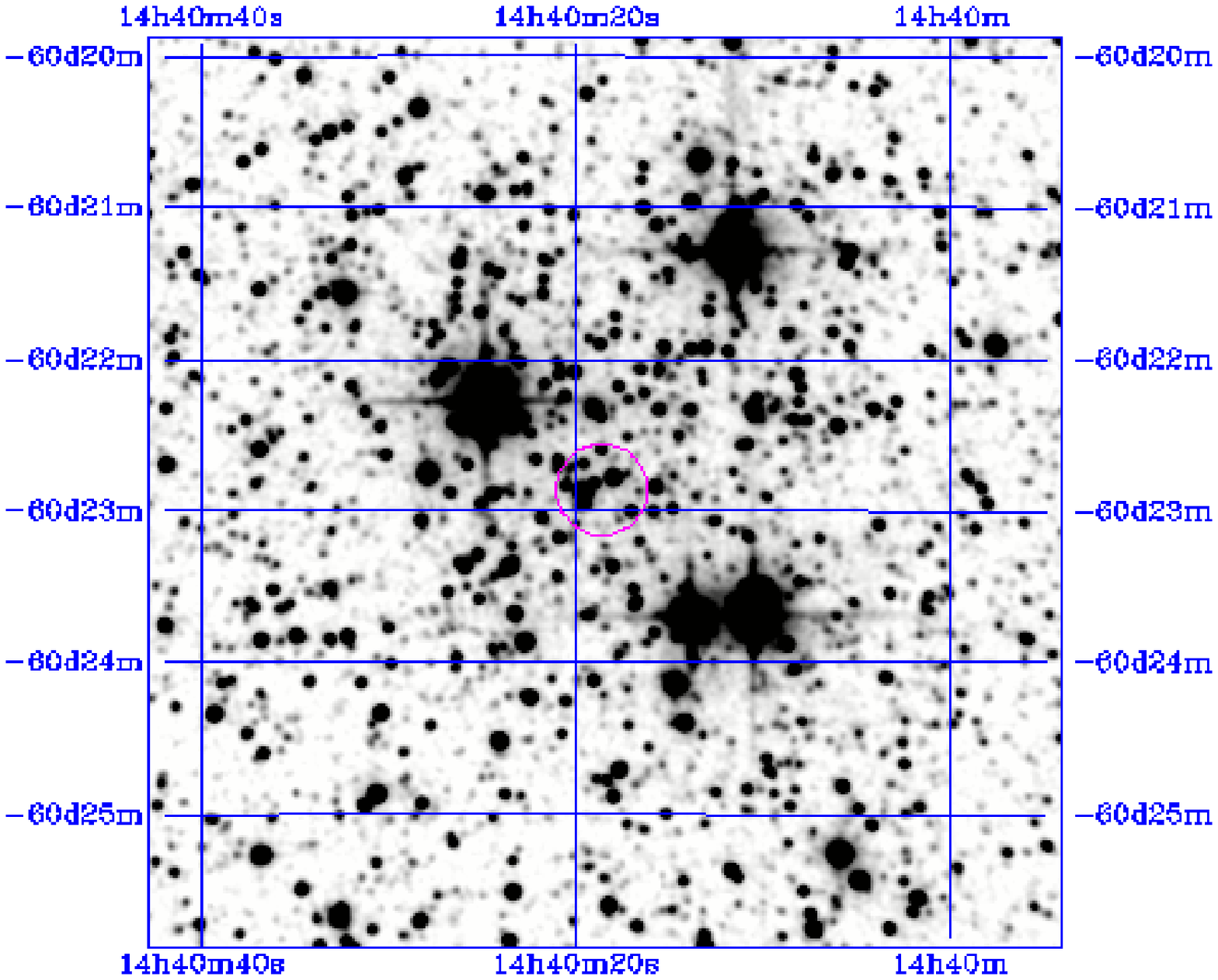}
\end{minipage}\hfill
\caption[]{2MASS \ks\ images of Mayer\,1 ($5\arcmin\times5\arcmin$), Muzzio\,1
($5\arcmin\times5\arcmin$), and Juchert\,10 ($6\arcmin\times6\arcmin$).}
\label{fig_APB1}
\end{figure*}

\label{lastpage}
\end{document}